\documentclass[12pt]{iopart}
\usepackage{amssymb}
\usepackage{graphics, graphicx}
\usepackage{subfigure}
\usepackage{color}
\usepackage{soul}

\def\beq{\begin{equation}}
\def\eeq{\end{equation}}
\def\bea{\begin{eqnarray}}
\def\eea{\end{eqnarray}}

\def\be{\begin{equation}}
\def\ee{\end{equation}}
\begin{document}

\title{Magnetism and domain formation in SU(3)-symmetric multi-species Fermi mixtures}
\author{I.~Titvinidze$^{1,5}$, A.~Privitera$^{1,2,5}$, S.-Y.~Chang$^{3,4}$, S.~Diehl$^{3}$, M.~A.~Baranov$^{3}$, A.~Daley$^{3}$ and W.~Hofstetter$^{1}$}

\address{$^{1}$ Institut f\"ur Theoretische Physik, Johann Wolfgang Goethe-Universit\"at, 60438 Frankfurt am Main, Germany}
\address{$^{2}$ Dipartimento di Fisica, Universit\`a di Roma La Sapienza, Piazzale Aldo Moro 2, 00185 Roma, Italy}
\address{$^{3}$ Institute for Quantum Optics and Quantum information of the Austrian Academy of Sciences,
                 A-6020 Innsbruck, Austria, Institute for Theoretical Physics, University of Innsbruck, A-6020 Innsbruck, Austria}
\address{$^{4}$ Department of Physics, The Ohio State University, Columbus, OH 43210, USA}
\address{$^{5}$ These authors contributed equally to this work}
\ead{irakli@itp.uni-frankfurt.de}
\date{\today}

\begin{abstract}
We study the phase diagram of an SU(3)-symmetric mixture of three-component ultracold fermions with attractive interactions in an optical lattice,
including the additional effect on the mixture of an effective three-body constraint induced by three-body losses. We address the 
properties of the system in $D \geq 2$ by using dynamical mean-field theory and variational Monte Carlo techniques. The phase diagram of
the model shows a strong interplay between magnetism and superfluidity. In the absence of the three-body constraint (no losses), the 
system undergoes a phase transition from a color superfluid phase to a trionic phase, which shows additional particle density modulations
at half-filling. Away from the particle-hole symmetric point the color superfluid phase is always spontaneously magnetized,
leading to the formation of different color superfluid domains in systems where the total number of particles of each species is
conserved. This can be seen as the SU(3) symmetric realization of a more general tendency to phase-separation in 
three-component Fermi mixtures. The three-body constraint strongly disfavors the trionic phase, stabilizing a
(fully magnetized) color superfluid also at strong coupling. With increasing temperature we observe a transition
to a non-magnetized $SU(3)$ Fermi liquid phase.   
\end{abstract}
\maketitle
\section{Introduction}

Cold atoms in optical lattices provide us with an excellent tool to investigate notoriously difficult problems in condensed matter
physics \cite{hof0,Bloch2008}. Recent progress towards this goal is exemplified by the experimental observation of the fermionic 
Mott insulator \cite{Hubbard1,Hubbard2} in a binary mixture of repulsively interacting $^{40}{\rm K}$ atoms loaded into an optical
lattice, and of the crossover between Bardeen-Cooper-Schrieffer (BCS) superfluidity and Bose-Einstein condensation
(BEC) \cite{superfluid_Ketterle, superfluid_Grimm, superfluid_Jin} in a mixture of $^6{\rm Li}$ atoms
with attractive interactions. 

At the same time, ultracold quantum gases also allow us to investigate systems which have no immediate counterparts
in condensed matter. This is the case for fermionic mixtures where three internal states $\sigma=1,2,3$ are used, instead of the
usual binary mixtures that mimic the electronic spin $\sigma=\uparrow, \downarrow$. These multi-species Fermi mixtures are already 
available in the laboratory, where three different magnetic sublevels of $^{6}{\rm Li}$ \cite{selim, Wenz09, Huckans09, Williams}
 or  $^{173}{\rm Yb}$ \cite{ytterbium}, as well as a mixture of the two internal states of $^6{\rm Li}$ with
a lowest hyperfine state of $^{40}{\rm K}$\cite{Wille} have been successfully trapped. In the case of Alkali atoms, magnetic or 
optical Fano-Feshbach resonances can be used to tune magnitude and sign of the interactions in the system, and in the case of
Ytterbium or group II atoms, it is possible to realise three-component mixtures where the components differ only by nuclear spin,
 and therefore exhibit SU(3) symmetric interactions \cite{Gorshkov, Takahashi,Cazalilla}. Moreover, loading these mixtures
into an optical lattice would give experimental access to intriguing physical scenarios, since they can realize 
a three-species Hubbard model with a high degree of control of the Hamiltonian parameters. 
 
Multi-species Hubbard models have attracted considerable interest on the theoretical side in recent years. First studies
were focused on the $SU(3)$-symmetric version of the model with attractive interaction. By using a generalized BCS approach
\cite{hof1, hof2, Leggett}, it was shown that the ground state at weak-coupling spontaneously breaks the $SU(3) \otimes U(1) $ 
symmetry down to $SU(2)\otimes U(1)$, giving rise to a color superfluid (c-SF) phase, where
superfluid pairs coexist with unpaired fermions. Within a variational Gutzwiller technique \cite{hof3,hof4}
 the superfluid phase was then found to undergo for increasing attraction a phase transition to a Fermi liquid
trionic phase, where bound states (trions) of the three different species are formed and the
$SU(3)$-symmetry is restored. More recently \cite{inaba, inaba2}, the same scenario has been found 
by using a self-energy functional approach for the half-filled
model on a Bethe lattice in dimension $D=\infty$. It was suggested \cite{QCD} that this transition  bears analogies to 
the transition between quark superfluid and baryonic phase in the context of Quantum Chromo Dynamics. 

Both the attractive and the repulsive version of the model was addressed by numerical and analytical techniques for the peculiar 
case of spatial dimension $D=1$ \cite{Molina,Azaria,kantian,Ulbricht}, while Mott physics and instabilities towards (colored) 
density wave formation have been found in the repulsive case in higher dimensions \cite{hof1,Gorelik,Miyatake}. It is important 
to mention that substantial differences are expected in the attractive case at strong coupling when the lattice is not present 
\cite{Paananen,flor}. Those differences are essentially related to the influence of the lattice in the strong coupling limit in the
three-body problem, favoring trion formation \cite{honerkamp,Jan} with respect to pair formation in the continuum, as was shown in 
Ref.~\cite{flor, Naidon, Braaten}.

Here we consider the $SU(3)$-symmetric system in a lattice for $D \geq 2$ in the presence of attractive two-body 
interactions by combining dynamical mean-field theory (DMFT) and variational Monte Carlo (VMC). We analyze several cases
of interest for commensurate and incommensurate density. Ground state, spectral, and finite temperature properties are 
addressed. More specifically we focus on the transition between color superfluid and trionic phase and on a better understanding of the 
coexistence of magnetism and superfluidity in the color superfluid phase already predicted in the $SU(3)$ symmetric case \cite{hof3,hof4}
but also when the $SU(3)$-symmetry is explicitly broken \cite{cherng}. We show that the existence of a spontaneous magnetization  
leads the system to separate in color superfluid domains with different realizations of color pairing and magnetizations
whenever the total number of particles in each hyperfine state is conserved. This would represent a special case, due to the
underlying $SU(3)$ symmetry, of a more general tendency towards phase separation in three-component Fermi mixtures. 
We point out that all this rich and interesting physics arises merely from having three components instead of two. 
Indeed the analogous $SU(2)$ system would give rise to the more conventional BCS-BEC crossover,
where the superfluid ground state evolves continuously for increasing attraction \cite{toschi}. Moreover in the $SU(2)$
case superfluidity directly competes with magnetism \cite{werner}. 

The case under investigation can be realized with ultracold gases by loading a three-species mixture
 of $^{173}{\rm Yb}$ \cite{ytterbium} or another group II element such as $^{87}$Sr into an optical lattice,
or alternatively using $^6{\rm Li}$ in a large magnetic field. However, some realizations with ultracold atoms are plagued by
three-body losses due to three-body Efimov resonances \cite{selim, Wenz09, Williams}, which are not any more Pauli suppressed 
as in the two-species case. The three-body loss properties and their dependence on the magnetic field 
have been already measured for $^6{\rm Li}$ \cite{selim, Wenz09, Williams}, while they are still unknown 
for three-component mixtures of certain group-II elements. Loading a gas into an optical lattice could be 
used to suppress losses, as a large rate of onsite three-body loss can prevent coherent tunneling processes
from populating any site with three particles \cite{daley}. As proposed in Ref. \cite{daley} for bosonic systems,
in the strong loss regime a Hamiltonian formulation is still possible if one includes an effective hard-core three-body
interaction, which leads to new interesting physics \cite{diehl1}. The effect of this dynamically generated constraint 
on the fermionic system in $D=1$ with attractive interactions was studied in Ref. \cite{kantian}, where it was shown that
the constraint may help to stabilize the superfluid phase in some regions of the phase diagram.   

For these reasons we also study the effect of including a three-body constraint 
in the model, as representative of an $SU(3)$ symmetric mixture in the strong-loss regime.  
The asymmetric case in the strong loss regime, which is directly relevant for experiments on 
$^6{\rm Li}$ close to a Feshbach resonance, has been already addressed in a separate publication \cite{asymm_short}.

The paper is organized as follows: in the following sections we first introduce the model (Sec. \ref{Model}) 
and then the methods used (Sec. \ref{Methods}). Later on we present our results, focusing first on the 
unconstrained system (Sec. \ref{unconstrained}), for commensurate and incommensurate densities and then
on the effects of the three-body constraint (Sec. \ref{constrained}). 
The emergence of domain formation within globally \emph{balanced} mixtures is discussed in detail
 in Sec. \ref{Phase_Separation}. Final remarks are drawn in Section \ref{conclusion}. 
 
\section{Model}\label{Model}

Three-component Fermi mixtures with attractive two-body interactions loaded into an optical lattice are 
well described by the following Hamiltonian
\begin{eqnarray}
\label{hamiltonian}
\hspace{-1.75cm}{\cal{H}}=-J\sum_{\langle i,j\rangle,\sigma}c^\dagger_{i,\sigma}c_{j,\sigma} -\sum_{i,\sigma}\mu_\sigma n_{i,\sigma} +\sum_i \sum_{\sigma <
\sigma^\prime} U_{\sigma\sigma^\prime} n_{i,\sigma}n_{i\sigma^\prime} +V\sum_i n_{1,i} n_{2,i} n_{3,i} \, ,
 \end{eqnarray}
where $\sigma=1,2,3$ denotes the different components, $J$ is the hopping parameter between nearest neighbor sites 
$\langle i,j\rangle$, $\mu_\sigma$ is the chemical potential for the species $\sigma$ and $U_{\sigma\sigma^\prime} < 0$.
We introduced the onsite density operators $n_{i\sigma} = c_{i\sigma}^\dag c_{i \sigma}$. The three-body interaction term with
$V=\infty$ is introduced to take the effects of three-body losses in the strong loss regime into
account according to Refs. \cite{kantian,daley}. $V=0$ corresponds to the case when three-body losses are
negligible. While the model and the methods are developed for the general case without $SU(3)$-symmetry,
in this paper we concentrate on the $SU(3)$-symmetric case reflected by species-independent parameters
\begin{eqnarray}
U_{\sigma\sigma^\prime}=U, \quad \mu_\sigma=\mu.
\end{eqnarray}
In this case the Hamiltonian (\ref{hamiltonian}) reduces to an $SU(3)$ attractive Hubbard model if $V=0$. 
Note that the three-body interaction term is a color singlet and thus does not break $SU(3)$ for any choice of $V$.
On the basis of previous works, the ground state of the unconstrained model is expected to be, at least in the weak coupling regime, a
color superfluid, i.e. a phase where the full $SU(3)\otimes U(1)$ symmetry of the Hamiltonian is 
spontaneously broken to $SU(2)\otimes U(1)$ \cite{hof1, hof2}. As shown in \cite{hof1, hof2}, it is always possible to find a
suitable gauge transformation such that pairing takes place only between two of the natural species $\sigma$, $\sigma'$
and in this paper we choose a gauge in which pairing takes place between the species $\sigma=1$ and $\sigma'=2$ ($1-2$ channel),
while the third species stays unpaired. Whenever the $SU(3)$-symmetry is
\emph{explicitly} broken, only the pairing between the natural species is allowed to comply with Ward-Takahashi identities
\cite{cherng}. This reduces the continuum set of equivalent pairing channels of the symmetric model to a discrete set of 
three (\emph{mutually exclusive}) options for pairing, i.e. $1-2,1-3$ or $2-3$. In this case the natural choice would be that pairing 
takes place in the channel corresponding to the strongest coupling when the mixture is globally balanced. We can always
relabel the species such that strongest attractive channel is the channel $1-2$. Other pairing channels can be studied via
index permutations of the species. Therefore the formalism developed here is fully general and includes both the
symmetric and non-symmetric case, while only in the $SU(3)$-symmetric case our approach
corresponds to a specific choice of the gauge.

\section{Methods}\label{Methods}
In order to investigate the model in Eq. (\ref{hamiltonian}) in spatial dimensions $D\geq2$ we use a combination of
numerical techniques which have proven to give very consistent results for the non-symmetric case  \cite{asymm_short}. In particular,
we use dynamical mean-field theory (DMFT) for $D \geq 3$ and variational Monte Carlo (VMC) for $D=2$. DMFT provides us with the 
exact solution in infinite dimension and a powerful (and non-perturbative) approach in $D=3$,
which has the advantage of being directly implemented in the thermodynamic limit (without finite size effects). VMC allows us 
to incorporate also the effect of spatial fluctuations which are not included within DMFT, even though the exponential growth 
of the Hilbert space limits the system sizes that are
accessible.
%  Finally the analytical approach provides us with an approximate but simpler point of view on the basic physical mechanism involved.      

\subsection{DMFT}\label{DMFT}   
Dynamical mean-field theory (DMFT) is a non-perturbative technique based on the original idea of Metzner and Vollhardt who studied the limit of infinite
dimension of the Hubbard model \cite{metzner}. In this limit, the self-energy $\Sigma(\bf{k},\omega)$ becomes momentum independent 
$\Sigma({\bf{k}},\omega)=\Sigma(\omega)$, while fully retaining its frequency dependence. Therefore the many-body problem simplifies
significantly, without becoming trivial, and can be solved exactly. In this sense DMFT is a \emph{quantum} version of the static 
mean-field theory for classical systems, since it becomes exact in the same limiting case ($D=\infty$) and can provide useful 
information also outside of this limit, fully including local quantum fluctuations. In 3D, assuming a
momentum independent self-energy, has proved to be a very accurate approximation for many problems where the momentum
dependence is not crucial to describe the physics of the system such as the Mott metal-insulator transition \cite{georges_rmp}
where the frequency dependence is more relevant than the $\bf{k}$ dependence.

\subsubsection{ Theoretical setup for $SU(3)$ model with spontaneous symmetry breaking --}\label{sec:thset} In this work, we generalize the DMFT
approach to multi-species Fermi mixtures in order to describe color superfluid and trionic phases, which are the expected phases occurring in the system. The theory
can be formulated in terms of a set of self-consistency equations for the components of the local single-particle Green function $\hat{G}$ on the lattice. Since we
are dealing here with superfluid phases involving also anomalous components of the Green function, we use a compact notation in terms of mixed Nambu spinors
$\psi=(c_1,c^\dagger_2,c_3)$, where we already assumed that pairing takes place only between the first two species, as explained in the previous section, and we omit
the subscript $i$ (spatially homogeneous solution). We reiterate that this specific choice is valid without loss of generality in the  $SU(3)$-symmetric
model, and has the same status as fixing the phase of a complex condensate order parameter in theories with global phase symmetry. The local Green function
$\hat{G}(i\omega_n)$ in Matsubara
space then has the form 
\begin{equation}
\label{Green_matrix_local2}
\hat G(i\omega_n)= \left(
\begin{array}{ccc}
G_1(i\omega_n)& F(i\omega_n) & 0\\
F^*(-i\omega_n) & -G_2^*(i\omega_n)& 0\\
0 & 0 & G_{3}(i\omega_n)
\end{array}
\right) \, ,
\end{equation}
where $G_\sigma(\tau)=-\langle T_\tau c_\sigma(\tau) c^\dagger_\sigma \rangle$ and 
$F(\tau)=-\langle T_\tau c_1(\tau) c_2 \rangle$ are respectively the normal and
anomalous Green functions in imaginary time, and $G_\sigma(i\omega_n)= \int_0^\beta d\tau G_\sigma (\tau) 
e^{i\omega_n\tau}$ and $ F(i\omega_n)=\int_0^\beta F(\tau)
e^{i\omega_n\tau}$ are their Fourier transforms in Matsubara space, where $\omega_n=(2n+1)\pi T$ ($k_B=1$). 

In practice the original lattice model (\ref{hamiltonian}) in the DMFT approach can be mapped, by introducing
auxiliary fermionic degrees of freedom $a_{l\sigma}^{\dagger},a_{l\sigma}^{\phantom\dagger}$, on a Single Impurity
Anderson Model (SIAM), whose Hamiltonian reads
\begin{eqnarray}
\label{hamilton_and}
{\cal H}_{SIAM}&=&\sum_{\sigma < \sigma^\prime} U_{\sigma\sigma^\prime} n_{\sigma}n_{\sigma^\prime}+V n_{1}n_{2} n_{3}-
\sum_{\sigma}\mu_\sigma n_{\sigma} \\
&&+\sum_{l\sigma}\left[\varepsilon_{l\sigma}a_{l\sigma}^{\dagger} a_{l\sigma}^{\phantom\dagger}+ 
V_{l\sigma} \left(c_{\sigma}^{\dagger}a_{l\sigma}^{\phantom\dagger}+h.c.\right)\right] 
+\sum_{l} W_{l}\left[a_{l,1}^{\dagger}a_{l,2}^{\dagger}+h.c.\right]  , \nonumber 
\end{eqnarray}
where the Anderson parameters $\varepsilon_{l\sigma},V_{l\sigma},W_{l}$ have to be determined self-consistently.
Self-consistency ensures that the impurity Green function of the SIAM is identical
to the local component of the lattice Green function. The components
of the non-interacting Green function for the impurity site, which represent the dynamical analog of the Weiss field in 
classical  statistical mechanics, can be expressed in terms of the Anderson parameters
as 
\begin{eqnarray}
\label{Weiss_Green_1}
&&{\cal G}^{-1}_{1,And}(i\omega_{n})=i\omega_{n}+\mu_1+\sum_{l=1}^{n_s} \frac{V_{l,1}^2 \zeta_{l,2}^*}{\zeta_{l,1}\zeta_{l,2}^*+W_l^2 } \, , \\
\label{Weiss_Green_2}
&&{\cal G}^{-1}_{2,And}(i\omega_{n})=i\omega_{n}+\mu_2+\sum_{l=1}^{n_s}\frac{V_{l,2}^2 \zeta_{l,1}^*}{\zeta_{l,2}\zeta_{l,1}^*+W_l^2 } \, ,\\
\label{Weiss_Green_SC}
&&{\cal F}^{-1}_{SC,And}(i\omega_{n})=\sum_{l=1}^{n_s}\frac{V_{l,1}V_{l,2} W_l}{\zeta_{l,1}\zeta_{l,2}^*+W_l^2} \, ,\\
\label{Weiss_Green_3}
&&{\cal G}^{-1}_{3,And}(i\omega_{n})=i\omega_{n}+\mu_3+\sum_{l=1}^{n_s}\frac{V_{l,3}^2}{\zeta_{l,3}} \, ,
\end{eqnarray}
where $\zeta_{l,\sigma}=-i\omega_n+\varepsilon_{l\sigma}$. The self-consistency equations for the local Green functions now have the form 
\begin{equation}
\label{self}
\hat{G}(i\omega_n)
=\frac{1}{M}\sum_{\bf{k}}\hat{G}^{latt}({\bf{k}},i\omega_n)=\int d\varepsilon D(\varepsilon)\hat{G}^{latt}(\varepsilon,i\omega_n) \, ,
\end{equation}
where $M$ is the number of lattice sites, $\hat{G}^{latt}({\bf{k}},i\omega_n)=\hat{G}^{latt}(\varepsilon_{\bf{k}},i\omega_n)$ is the
lattice Green function within DMFT and $D(\varepsilon)$ is the 
density of states of the lattice under consideration. The independent components of 
$\hat{G}^{latt}({\bf{k}},i\omega_n)$ have the form 
\begin{eqnarray}
\label{Glatt1}
\hspace{-0.6cm}G^{latt}_1=
\!\!&&\frac{\zeta_2^*-\varepsilon_{\bf{k}}}{(\zeta_1-\varepsilon_{\bf{k}})(\zeta^*_2-\varepsilon_{\bf{k}})+\Sigma_{SC}(i\omega_n)\Sigma_{SC}^*(-i\omega_n)} \, ,\\ 
\label{Glatt2}
\hspace{-0.6cm}G^{latt}_2=
\!\!&&\frac{\zeta_1^*-\varepsilon_{\bf{k}}}{(\zeta_2-\varepsilon_{\bf{k}})(\zeta^*_1-\varepsilon_{\bf{k}})+\Sigma_{SC}(-i\omega_n)\Sigma_{SC}^*(i\omega_n)} \, ,\\ 
\label{Flatt}
\hspace{-0.6cm}F^{latt}_{SC}=
\!\!&&-\frac{\Sigma_{SC}(i\omega_n)}{(\zeta_1-\varepsilon_{\bf{k}})(\zeta^*_2-\varepsilon_{\bf{k}})+\Sigma_{SC}^2(i\omega_n)}  \, ,\\
\label{Glatt3}
\hspace{-0.6cm}G^{latt}_{3}=
\!\!&& \frac{1}{\zeta_3-\varepsilon_{\bf{k}}} \, ,
\end{eqnarray}
where $\zeta_\sigma=i\omega_n+\mu_\sigma-\Sigma_\sigma(i\omega_n)$ and the self-energy can be obtained by the 
following local Dyson equation ${\hat\Sigma}(i\omega_n)={\hat{\cal G}}^{-1}_{And}(i\omega_n)-{\hat G}^{-1}(i\omega_n)$
where
\begin{equation}
\label{Self_energy}
\hat \Sigma(i\omega_n)= \left(
\begin{array}{ccc}
\Sigma_1(i\omega_n)& \Sigma_{SC}(i\omega_n)& 0\\
\Sigma_{SC}^*(-i\omega_n) & -\Sigma_2^*(i\omega_n)& 0\\
0 & 0 & \Sigma_3(i\omega_n)
\end{array}
\right) \, .
\end{equation}

Once a self-consistent solution has been obtained, the impurity site of the SIAM represents a generic site of the lattice model under investigation. Therefore
several static thermodynamic quantities can be directly evaluated as quantum averages of the impurity site. As evident from the previous equations, DMFT is
explicitly formulated in a grand canonical approach where the chemical potentials $\mu_\sigma$ are given as input and the onsite densities $n_\sigma=\langle
c^\dagger_\sigma c_\sigma \rangle$ are calculated. 

\subsubsection{ Calculated observables and numerical implementation -- } To characterize the different phases, we evaluated several static observables
such as the superfluid (SF) order parameter $P=\langle c_1 c_2 \rangle$, the average double occupancy $d_{\sigma\sigma^\prime}=\langle n_\sigma
n_{\sigma^\prime}\rangle$ and the average triple occupancy $t=\langle n_1 n_2 n_3\rangle$. As suggested in Refs. \cite{hof1, hof2, cherng}, in order to gain
condensation energy in the c-SF phase, it is energetically favorable to induce a finite density imbalance between
the paired species ($1-2$ in our gauge) and the unpaired fermions. To \emph{quantitatively} characterize this feature we 
introduce the local \emph{magnetization}
\begin{equation}
\label{magnetization}
 m=n_{12} -n_3 \ \ \mbox{where}\ \  n_{12}=n_1=n_2  \, .
\end{equation}

From the normal components of the lattice Green functions in Eqs. (\ref{Glatt1}), (\ref{Glatt2}) and (\ref{Glatt3})
we can extract the DMFT momentum distribution
\begin{equation}
 n_\sigma({\bf{k}})=T\sum_n G^{latt}_\sigma({\bf{k}},i\omega_n)e^{-i\omega_n0^-}
\end{equation}
 and the average of kinetic energy per lattice site
\begin{equation}
K=\frac{1}{M}\sum_{{\bf{k}},\sigma}\varepsilon_{\bf{k}} n_\sigma({\bf{k}})=\sum_\sigma\int d\varepsilon \ 
D(\varepsilon)\  \varepsilon\  n_\sigma(\varepsilon).
\end{equation}
 It is evident from the expression of  $\hat{G}^{latt}({\bf{k}},i\omega_n)$ given in Eqs. 
(\ref{Glatt1}), (\ref{Glatt2}) and (\ref{Glatt3}) that
$n_\sigma({\bf{k}})$ only depends on the momentum $\bf{k}$ through the free-particle dispersion
$\varepsilon_{\bf{k}}$ of the lattice at hand. The internal energy per lattice site $E$ can then be
obtained as 
\begin{equation}
 E=K+V_{pot},\ \ \mbox{where}\ \ V_{pot}=\sum_{\sigma \neq \sigma^\prime} \frac{U_{\sigma \sigma^\prime}}{2} d_{\sigma\sigma^\prime}
\end{equation}
is the average potential energy per lattice site. 

Solving the DMFT equations is equivalent to solving a SIAM in presence of a bath determined self-consistently. We use Exact Diagonalization (ED) \cite{krauth},
which amounts to truncating the number of auxiliary degrees of freedom $a_{l\sigma}^{\dagger},~a_{l\sigma}^{\phantom\dagger}$ in the Anderson model to a
finite (and small) number $N_s-1$. In this way the size of the Hilbert space of the SIAM is manageable and we can exactly solve the Anderson model numerically. Here
we would like to point out that this truncation does not reflect the size of the physical lattice but only the number of independent parameters
used in the description of the local dynamics. Therefore we always describe the system in the thermodynamic limit (no finite-size effects). We use the Lanczos
algorithm \cite{matr_comp} to study the ground state properties (up to $N_s=7$) and full ED for finite temperature (up to $N_s=5$). Due to the increasing size of the
Hilbert space ($\sigma=1,2,3$ instead of $\sigma=\uparrow,\downarrow$) in the multi-component case the typical values of $N_s$ which can be handled sensibly is
smaller than the corresponding values for the $SU(2)$ superfluid case. However, in thermodynamic quantities, we found indeed only a very weak dependence on the value
of $N_s$ and the results within full ED at the lowest temperatures are in close agreement with $T=0$ calculations within Lanczos.

A definite advantage of ED is that it allows us to directly calculate dynamical observables for real frequencies without need of analytical
continuation from imaginary time. In particular, we can directly extract the local single-particle Green function  $G_\sigma(\omega)$ and the single-particle
spectral function
\begin{equation}
\rho_\sigma(\omega)=-\frac{1}{\pi} Im G_\sigma(\omega+i0^+).
\end{equation}

\subsection{Variational Monte-Carlo}\label{VMC}

The variational Monte Carlo (VMC) techniques described in this subsection can be used to calculate the energies and correlation functions of the homogeneous
phases at $T=0$ in a canonical framework. The basic ingredients of the VMC formalism are the Hamiltonian and trial wavefunctions with an appropriate symmetry.  In
principle, the formalism presented here can be applied to any dimension, even though here we use it specifically to address the system on a two-dimensional square
lattice.

The canonical version of Hamiltonian (\ref{hamiltonian}) for three-components fermions with generic attractive interactions is given by 
\begin{equation}
{\cal H}  =  -J \sum\limits_{\langle i,j \rangle,\sigma} {\cal P}_3 c^\dagger_{i,\sigma} c_{j,\sigma} {\cal P}_3 + \sum\limits_{i,\sigma<\sigma'}
U_{\sigma\sigma^\prime} n_{i,\sigma} n_{i,\sigma'}  ,
\label{eqn_hbare}
\end{equation}
where the three-body constraint is imposed by using the projector ${\cal P}_3 = \prod_i (1-n_{i,1} n_{i,2} n_{i,3})$ and in the unconstrained case
we set ${\cal P}_3$ equal to the identity. 

Practical limitations do not permit a general trial wave function equally accurate both for the weak- and the strong-coupling limit. Due to this reason
we introduce different trial wavefunctions for different coupling regimes.  

In the weakly interacting limit, which we operatively define as $|U_{\sigma\sigma^\prime}| \leq 4J=W/2$, we use the full Hamiltonian (\ref{eqn_hbare}) along with the
weak-coupling trial wavefunction defined in the next subsection. Here $W = 2DJ$ is the bandwidth. At strong-coupling this wavefunction results in a poor description of the
system. In order to gain insight into the strong coupling regime, we derive below a perturbative Hamiltonian to the second order in $J/U_{\sigma\sigma^\prime}$,
which we will combine with a strong-coupling trial wavefunction. Again the strong-coupling wavefunctions are incompatible with the Hamiltonian (\ref{eqn_hbare}), as
will be clarified below. We can therefore address confidently both limits of the model while at intermediate coupling we expect our VMC results to be less accurate.

\subsubsection{Strong coupling Hamiltonian,  constrained case  --}\label{VMC_strong_coupling_H}

In order to derive a perturbative strong-coupling Hamiltonian for the constrained case we make use of the Wolff-Schrieffer transformation
\cite{macdonald88}
\begin{equation}
{\cal H}_{pert} = {\cal P}_D e^{i{\cal S}} {\cal H} e^{-i{\cal S}} {\cal P}_D 
\end{equation}
and keep terms up to the second order in $J/U_{\sigma\sigma^\prime}$. In the expression above, ${\cal P}_D$ is the projection operator to the Hilbert subspace with
fixed numbers of double occupancies in each channel ($N_{d}^{12}$, $N_{d}^{23}$,$N_{d}^{13}$), and $e^{i{\cal S}}$ is a unitary transformation defined in
\ref{Appendix}.  So  we obtain the perturbative Hamiltonian (see \ref{Appendix}), which reads:
\begin{eqnarray}
{\cal H}_{pert}  & = &  -J\sum\limits_{\langle i,j \rangle\sigma} f_{i,\sigma}^\dagger f_{j,\sigma} 
- J^2 \sum\limits_{\langle j',i \rangle; \langle i,j \rangle;\sigma < \sigma'} \frac{1}{U_{\sigma\sigma^\prime}} d^\dagger_{j',\sigma\sigma'} f_{i,\sigma}
f^\dagger_{i,\sigma} d_{j,\sigma\sigma'} \nonumber \\
&-& J^2 \sum\limits_{\langle i,j' \rangle; \langle i,j \rangle;\sigma < \sigma'} \frac{1}{U_{\sigma\sigma^\prime}} d^\dagger_{i,\sigma'\sigma} f_{j',\sigma'}
f^\dagger_{i,\sigma} d_{j,\sigma\sigma'} 
+ {\cal V} + {\cal O}(\frac{J^3}{U_{\sigma \sigma'}^2}) \, ,
\label{eqn_pert1}
\end{eqnarray}
where ${\cal V} = \sum\limits_{i,\sigma<\sigma'} U_{\sigma\sigma^\prime} n_{i,\sigma} n_{i,\sigma'}$. Here we define double occupancy operators as
$d_{i,\sigma\sigma'}^\dagger \equiv c_{i,\sigma}^\dagger n_{i,\sigma'}  h_{i,\sigma''}$  ($h_{i,\sigma} = 1 - n_{i,\sigma}$) and single occupancy operators
as $f_{i,\sigma}^\dagger \equiv h_{i,\sigma'}h_{i,\sigma''} c_{i,\sigma}^\dagger$ .

For the case where the $SU(3)$-symmetry is restored ($U_{\sigma\sigma^\prime}=U$), the perturbative Hamiltonian can be written in a compact notation 
\begin{eqnarray}
\label{eqn_pert2}
&&\hspace{-2cm}{\cal H}_{pert} ={\cal V} -J\sum\limits_{\langle i,j \rangle\sigma} \left[ f_{i,\sigma}^\dagger f_{j,\sigma} +  d_{i,\sigma}^\dagger d_{j,\sigma}
\right] 
- \frac{J^2}{U} \sum\limits_{\langle i',i \rangle; \langle i,j \rangle;\sigma} d^\dagger_{i',\sigma} f_{i,\sigma} f^\dagger_{i,\sigma} d_{j,\sigma} \\
&& \hspace{-1.25cm}
- \frac{J^2}{U} \sum\limits_{\langle i,j' \rangle; \langle i,j \rangle;\sigma'\ne\sigma} d^\dagger_{i,\sigma'} f_{j',\sigma'} f^\dagger_{i,\sigma} d_{j,\sigma} 
+ \frac{J^2}{U} \sum\limits_{\langle i',i \rangle\sigma'; \langle i,j \rangle\sigma} f^\dagger_{i',\sigma'} d_{i,\sigma'} d^\dagger_{i,\sigma} f_{j,\sigma} 
+  {\cal O}(\frac{J^3}{U^2}) \nonumber
\end{eqnarray}
where the double occupancy operator is now defined as $d_{i,\sigma}^\dagger = c_{i,\sigma}^\dagger  (h_{i,\sigma'}  n_{i,\sigma''}
+h_{i,\sigma''}  n_{i,\sigma'})$.

Now, rather than conserving the number of double occupancies $N_d^{\sigma\sigma'}$ in each channel, only the total number $N_{d,0}=N_d^{12}+N_d^{13}+N_d^{23}$  is
conserved due to the $SU(3)$-symmetry. Indeed Eq. (\ref{eqn_pert2}) contains terms where the tightly bound dimers are allowed to change the composition through 
second order processes. Thus, the $SU(3)$-symmetric case, in contrast to the case with strongly anisotropic interactions is qualitatively different from the
Bose-Fermi mixture, because the bosons - tightly bound dimers - can change composition as described above, while such a process was not allowed in the case of
the strong anisotropic interactions. We also notice that the last of the $\sim J^2/U$ terms contributes only when $N_{d,0} < N/2$. 

 \subsubsection{Strong coupling Hamiltonian, unconstrained case -- } Without the 3-body constraint  three fermions with different hyperfine states can occupy the same lattice site and we expect
them to form trionic bound states at sufficiently strong coupling. Correspondingly the many-body system should be in a trionic phase with heavy trionic
quasiparticles, as mentioned in previous studies
\cite{hof1,hof2,inaba,inaba2}. Therefore we expect that our perturbative approach can provide a description of the trions in the strong coupling limit.
%strong-coupling Hamiltonian describes in this case the behavior the trions.  
 
First we consider the extreme case $J=0$. In this limit formation of local trions takes place, i.e each site is either empty or occupied by three fermions with
different hyperfine spins. Their spatial distribution is random, because any distribution of trions will have the same energy. 
For finite $J$ with
$J\ll |U_{\sigma,\sigma'}|$ the hopping term can break a local trion, but this would result in a large energy penalty.

According to perturbation theory up to third order we could have two different contributions:  (i) one of the fermions hops to one of the neighboring sites and
returns back to the original site (second order perturbation), (ii) all three  fermions hop to the same nearest neighbor site (third order perturbation).  As
we show below, due to the first process there is an effective interaction between trions on nearest neighbor sites. Also due to this process the onsite energy
has to be renormalized. The second process (ii) describes the hopping of a local trions to a neighboring site.

After straightfroward calculations (see \ref{Appendix}) we obtain that the effective interaction between two trions on neighboring sites is 
\begin{equation}
V_{eff}=\Delta E_1-\Delta E_0= -\left(\frac{J^2}{U_{12}+U_{13}}+ \frac{J^2}{U_{12}+U_{23}}+ \frac{J^2}{U_{13}+U_{23}}\right) \, .
\end{equation}
For the $SU(3)$-symmetric case this expression is simplified and we obtain
\begin{equation}
V_{eff}= -\frac{3J^2}{2U}=\frac{3J^2}{2|U|} \,.
\end{equation}
Therefore the nearest-neighbour interaction between trions is repulsive in the $SU(3)$-symmetric case. 

For the hopping coefficient we obtain
\begin{equation}
J_{eff}=\sum_{\sigma,\sigma^\prime}^{\sigma \not=\sigma^\prime}\frac{J^3}{(U_{\sigma\sigma'}+U_{\sigma\sigma''}) (U_{\sigma\sigma''}+U_{\sigma'\sigma''})} 
\end{equation}
where $\sigma$, $\sigma'$ and $\sigma''$ are different from each other in the sum.

In the $SU(3)$-symmetric case, the expression again simplifies to
\begin{equation}
J_{eff}=\frac{3J^3}{2U^2}
\end{equation}

So we obtain the following effective Hamiltonian \cite{csaba}
\begin{equation}
\label{trion_Hamiltonian}
{\cal H}_{eff}=-J_{eff} \sum_{\langle i,j\rangle}t_i^\dagger t_j + 
V_{eff}\sum_{\langle i,j\rangle} n_i^T n_j^T \, .
\end{equation}
Here $t_i^\dagger$ is the creation operator of a local trion at lattice site $i$ and $n_{i}^T =t_i^\dagger t_i$ is the trionic
 number operator. Because the effective hopping of trions results from a third order process and the interaction from second order,
more precisely $J_{eff} = J/|U| \cdot V_{eff}$, the
effective trion theory is interaction dominated. Since the interaction describes nearest-neighbour repulsion, the strong coupling limit clearly favors a checkerboard
charge density wave ground state at half-filling 
\footnote{Despite our fermions are charge neutral, we use sometimes the expression charge density wave in analogy with the terminology commonly used in condensed matter physics.}, which we will discuss in more detail in Sec. \ref{unconstrained}.

\subsubsection{Trial wavefunctions:}

In order to describe a normal Fermi liquid phase without superfluid pairing,
 we use the following trial wavefunction
\begin{equation}
|NFF \rangle = {\cal J}{\cal P}_3 {\cal P}_D \prod_\sigma \prod\limits_{\varepsilon_{\bf k,\sigma} \le \varepsilon_{F,\sigma}} 
c_{{\bf k},\sigma}^\dagger |0\rangle,
\label{eqn_su3}
\end{equation}
where $|0\rangle$ is the vacuum state and $\varepsilon_{{\bf k},\sigma} = -2 J (\cos(k_x) + \cos(k_y))$ for a 2D square lattice with only nearest-neighbor hopping. 
The dependence on the densities is included in the value of the non-interacting Fermi energy $\varepsilon_{F,\sigma}$. The wavefunction above has
no variational parameters except for the choice of Jastrow factor 
\begin{equation}
\hspace{-2.5cm}{\cal J}= \left\{
\begin{array}{l}
\exp(\nu_3\sum_i n_{i,1}n_{i,2}n_{i,3}) \quad\quad {\rm unconstrained~case,~weak~coupling}\\
\exp(\nu_{c}\sum_{\langle i,j \rangle}(n_{i,1}n_{i,2}n_{j,3}+n_{i,1}n_{i,3}n_{j,2}+n_{i,2}n_{i,3}n_{j,1}  ))\quad\quad {\rm constrained~case}
\end{array}
\right.  \, , 
\end{equation}
which takes into account the effect of the interaction. Here $\nu_3$ and $\nu_c$ are variational parameters and $\sum_{\langle i,j \rangle}$ is summation with
nearest neigbours. The weak-coupling version of the wavefunctions presented in this part is obtained by setting ${\cal P}_D$ equal to unity. 

We also consider the broken symmetry $SU(2)\otimes U(1)$ phase with s-wave pairing in the $1-2$ channel, whose trial wavefunction is given by
\begin{equation}
|c-SF \rangle  =  {\cal J}{\cal P}_3 {\cal P}_D \prod_{{\bf k}}\left[u_{\bf k} + v_{\bf k} c_{-{\bf k},1}^\dagger c_{{\bf k},2}^\dagger \right] 
 \prod_{\varepsilon_{{\bf k}',3}<\varepsilon_{F,3}}c_{{\bf k}',3}^\dagger|0\rangle \, ,
\label{eqn_su2u1}
\end{equation}
where $u_{\bf k}^2 = \frac{1}{2}\left( 1+(\epsilon_{\bf k} - \tilde \mu)/\sqrt{(\epsilon_{\bf k} - \tilde \mu)^2 +
(\Delta^{s}({\bf k}))^2} \right)$ and $v_{\bf k}^2 = 1 - u_{\bf k}^2$. In this case, in addition to the Jastrow factor $\cal J$, we have $\tilde \mu$ 
and $\Delta_0 $ as additional variational parameters. The s-wave gap function $\Delta^s({\bf k}) = \Delta_0$ has no $\bf k$ dependence. This parametrization of
$\Delta^s({\bf k})$ leads upon Fourier transform to a singlet symmetric pairing orbital $\phi^s({\bf r}_1, {\bf r}_2) = \phi^s({\bf r}_2,  {\bf r}_1)$.

In practice the optimization parameter $\Delta_0$ depends on the density $n$ as well as on the coupling strength $U$. Also, even at the same coupling strength
$U$, the $\Delta_0$ can be qualitatively different for the weak and the strong coupling ansatz (in the intermediate regime  $U \approx -5J$). On the other hand,
the parameter $\tilde \mu$  depends mostly on $n$ (and only weakly on $U$). The general tendency we observe is that $\Delta_0$ is suppressed beyond the
filling density $n \gtrsim 1$ in the presence of the constraint. Within a BCS mean-field theory approach,  the condensation energy $E_{cond}$ is easily related to
the order parameter $\Delta_0$, being $E_{cond} \propto \Delta_0^2$. We however calculate it explicitly from the definition by comparing the ground state energies of
the normal and the superfluid phases for the same density. Therefore we define   
\begin{equation}
E_{cond} = E_{NFF} - E_{c-SF} \, ,
\end{equation}
where
\begin{eqnarray}
&&E_{NFF} = \langle NFF| {\cal H} |NFF \rangle/\langle NFF|NFF \rangle \, ,\\
&&E_{c-SF} = \langle c-SF| {\cal H} |c-SF \rangle/\langle c-SF|c-SF \rangle \, . 
\end{eqnarray}
We also calculate the order parameter $P$ that characterizes the superfluid correlation by considering the long range behavior of the pair correlation function
\begin{equation}
P \equiv \lim\limits_{r \rightarrow \infty}  P(r) \equiv \sqrt{\frac{2}{M}  \sum\limits_{j} \langle B^\dagger_{j+r} B_{j} \rangle},
\end{equation}
where $B^\dagger_i \equiv c^\dagger_{i,1} c^\dagger_{i,2}$ and $M$ is the total number of the lattice sites.

Finally, in order to describe the trionic Fermi liquid phase we can use the following trial wavefunction
\begin{equation}
|Trion \rangle = {\cal J}_t \prod\limits_{\varepsilon_{\bf k} \le \varepsilon_{F}} t_{{\bf k}}^\dagger |0\rangle,
\label{eqn_trion}
\end{equation}
In this case the Jastrow factor
\begin{equation}
{\cal J}_t=\sum \exp(- \nu_t \sum_{<i,j>} n_i^T n_j^T) \, .
\end{equation}
Here $\nu_t$ is a variational parameter and  $\sum_{\langle i,j \rangle}$ is summation over nearest neigbours.

\section{Results: $SU(3)$ attractive Hubbard Model}
\label{unconstrained} 
We first consider the $SU(3)$ attractive Hubbard model described by the Hamiltonian (\ref{hamiltonian}) with $V=0$. 
In a physical realization with ultracold gases in optical lattices, this corresponds to a situation where three-body
losses are negligible. In order to address the effects of dimensionality and of particle-hole symmetry, we analyze 
several cases of interest, namely (i) an infinite-dimensional Bethe lattice in the commensurate case (half-filling),
(ii) a three-dimensional cubic lattice and (iii) a two-dimensional square lattice, the latter two in the incommensurate case.
In order to simplify comparison of results on different dimensions, 
we rescaled everywhere the energies by the bandwidth $W$ of the specific lattice under consideration. For a Bethe lattice 
in $D=\infty$ the bandwidth is related to the hopping parameter by $W=4J$, while for a $D$ dimensional hypercubic
lattice it is $W=4DJ$.

\subsection{Bethe lattice at half-filling}

We first consider the infinite dimensional case, for which the DMFT approach provides the \emph{exact} solution of the 
many-body problem whenever the symmetry breaking pattern of the system can be correctly anticipated.
For technical reasons we consider here the Bethe lattice in $D=\infty$, which has a well defined semicircular density of states,
given by the following expression
\begin{equation}
 D(\epsilon)=\frac{8}{\pi W^2}\sqrt{(W/2)^2-\epsilon^2}
\end{equation}
The simple form of the self-consistency relation for DMFT on the Bethe lattice introduces technical advantages, as explained below.
Moreover, we can directly compare our results with recent calculations for the same system within a 
Self-energy Functional Approach (SFA)\cite{inaba, inaba2}. 

In the absence of three-body repulsion, the Hamiltonian (\ref{hamiltonian}) is particle hole-symmetric
whenever we choose $\mu=U$. In this case the system is half-filled, i.e. $n_{\sigma}=\frac{1}{2}$ for 
all of $\sigma$ and $n=\sum_\sigma n_{\sigma}=1.5$. 

\begin{figure}
\begin{center}
\includegraphics[scale=0.275]{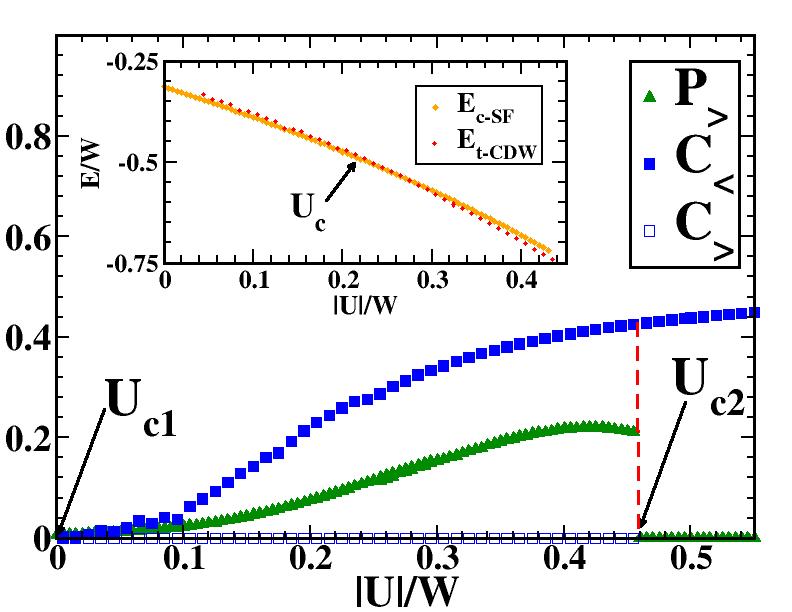}  
\end{center}
\vspace{-0.4cm}
\caption{(Color online) SF order parameter $P$ (green triangles) and CDW order parameter $C$ (blue squares) plotted as a function of interaction strength
$|U|/W$ on the Bethe lattice in the limit $D\rightarrow \infty$ at half-filling and $T=0$.  $C_>$ (empty squares) and $C_<$
(full squares) correspond to calculations  starting from a superfluid or a trionic charge density (t-CDW) initial conditions, and similar for $P_>$ (note that $P_<$
is always vanishing and therefore not shown). In the inset we compare the ground state energies of the c-SF and t-CDW phases. (Unconstrained, i.e. $V=0$)}
\label{static}
\end{figure}

We first consider the ground state properties of the system which we characterize via the static and dynamic
observables defined in Sec. \ref{Methods}. For small values of the interaction ($|U|\ll W$), we found the system
to be in a color-SF phase, i.e.  a phase where superfluid pairs coexist with unpaired fermions (species
1-2 and 3 respectively in our gauge) and the superfluid order parameter $P$ (plotted in Fig.~\ref{static} 
using green triangles) is finite. This result is in agreement with previous mean field 
studies \cite{hof1,hof2}, as expected since DMFT includes the (static) mean-field approach as a special limit,
 and with more recent SFA results \cite{inaba, inaba2}. 
By increasing the interaction $|U|$ in the c-SF phase, $P$ first increases continuously from a BCS-type 
exponential behavior at weak-coupling to a non-BCS regime at intermediate coupling where it shows a 
maximum and then starts decreasing for larger values of $|U|$. This non-monotonic behavior is beyond reach of a static
mean-field approach and agrees perfectly with the SFA result \cite{inaba,inaba2}. 
As explained in the introduction, the spontaneous symmetry breaking in the c-SF phase is
generally expected \cite{hof3,hof4,cherng,baym} to induce a population imbalance between 
the paired channel and the unpaired fermions, i.e. a finite value of the
magnetization $m$ in Eq. (\ref{magnetization}). It is however worth pointing out that,
 due to particle-hole symmetry, the c-SF phase at half-filling does not show any
induced population imbalance, i.e. $m=0$ for all values of the interaction strength.
As discussed in the next subsection, the population imbalance is indeed
triggered by the condensation energy gain in the paired channel. This energy gain, however,
cannot be realized at half-filling where the condensation energy is
already maximal for a given $U$.

%%%%%%%%%%%%%%%%%%%%%%%%%%%%%%%%%%%%%%%%%%%%%%%%%%%%%%%%%%%%%%%%%%%%%%%%%%%%%%%%%%%%%%%%%%%%%%%%%%%%%%%%  
%%%%%%%%%%%%%%%%%%%%%%%%%%%%%%%%%%%%%%%%%%%%%%%%%%%%%%%%%%%%%%%%%%%%%%%%%%%%%%%%%%%%%%%%%%%%%%%%%%%%%%%%

Further increasing $|U|$, we found $P$ to suddenly drop to zero at $|U|=U_{c,2} \approx 0.45 W$, signaling a first order transition 
to a non-superfluid phase. This result is in good quantitative agreement with the SFA result in Ref. \cite{inaba,inaba2}, 
where a first order transition to a trionic phase was found, while a previous
variational calculation found a second order transition \cite{hof3,hof4}.

In this new phase we were not able to stabilize a homogeneous solution of the DMFT equations with the ED algorithm \cite{krauth}. Such a spatially homogeneus phase
would correspond to having identical solutions, within the required tolerance, at iteration $n$ and $n+1$ of the DMFT self-consistency loop. In the normal phase
($|U| > U_{c,2}$) instead, we found a staggered pattern in the solutions and convergence is achieved if one applies a staggered criterion of convergence by comparing
the solutions in iteration $n$ and $n+2$. This behavior is clearly signalling that the transition to a non-superfluid phase is accompanied by a spontaneous symmetry
breaking of the lattice translational symmetry into two inequivalent sublattices A and B. In a generic lattice a proper description of this phase would require
solving two coupled impurity problems, i.e. one for each sublattice, and generalizing the DMFT equations introduced in the previous section. In the Bethe lattice
instead the two procedures are equivalent \footnote{On the Bethe lattice the sublattices A and B are completely decoupled from each other at a given step $n$.}.

In the new phase the full $SU(3)$-symmetry of the hamiltonian is restored and we identify it with as a trionic Charge Density Wave (t-CDW) phase. In order to
characterize this phase, we introduce a new order parameter which measures the density imbalance with respect to the sublattices A (majority) and B (minority), i.e.
\begin{equation}
C=\frac{1}{2}|n_A-n_B|
\end{equation}
where $n_A \equiv n_{\sigma,A}$ and $n_b \equiv  n_{\sigma,B}$ for all $\sigma$ and $C=0$ in the c-SF phase
because the translational invariance is preserved. 
The evolution of the CDW order parameter $C$ in the t-CDW is shown in Fig.~\ref{static} using blue squares. At the phase transition from c-SF to t-CDW phase, $P$
goes to zero and $C$ jumps from zero to a finite value. Then $C$ increases further with increasing attraction $|U|$ and eventually saturates at $C=1/2$ for 
$|U|\to\infty$. Motivated by these findings, we considered more carefully the region around the transition point. Surprisingly we found that upon decreasing $|U|$
from strong- to weak-coupling the t-CDW phase survives far below $U_{c,2}$ down to a lower critical value $U_{c1} \simeq 0$, revealing the existence of
a coexistence region in analogy with the hysteretic behavior found at the Mott transition in the single band Hubbard model \cite{georges_rmp}. In the present case,
however, we did not find any simple argument to understand which phase is stable and had to directly compare the ground state energy of the two phases in the
coexistence region to find the actual transition point. In the Bethe lattice, the kinetic energy per lattice site $K$ in the c-SF and t-CDW phases can be expressed
directly in terms of the components of the local Green function $\hat{G}(i\omega_n)$, which is straightforwardly determined by DMFT. The potential energy per lattice
site $V$ is given by $V_{t-CDW}= \frac{U(d_A+d_B)}{2}$, where the index indicates the sublattice. By generalizing analogous expressions valid in the $SU(2)$ case
\cite{toschi,dao, dao2}, we obtain 
\begin{equation}
\label{KcSF}
K_{c-SF} = T\sum_{n}(W/4)^2[\sum_\sigma G_\sigma^2(i\omega_n)-F^2(i\omega_n)]
\end{equation}
and
\begin{equation}
\label{KtCDW}
K_{t-CDW} = T\sum_{n,\sigma} (W/4)^2[G_A(i\omega_n) G_B(i\omega_n)].
\end{equation}
Results shown in the inset of Fig. \ref{static} indicate that the t-CDW phase is stable in a large part of the coexistence region and that the actual phase
transition takes place at $|U|=U_{c}\approx 0.2 W$. The good agreement between our findings and the SFA results in Ref.~\cite{inaba,inaba2} concerning
the maximum value of the attraction $U_{c2}$  where a c-SF phase solution is found within DMFT would suggest that this value is indeed a critical threshold for the
existence of a c-SF phase. On the other hand we also proved that the c-SF phase close to $U_{c2}$ is metastable with respect to the t-CDW phase and therefore the
existence of the threshold could equally results from an inability of our DMFT solver to further follow the metastable c-SF phase at strong coupling. The
disagreement between our findings and Ref.~\cite{inaba,inaba2} for what concerns the existence of CDW modulations in the trionic phase is clearly due to the
constraint of homogeneity imposed in the SFA approach of Ref. \cite{inaba,inaba2} in order to stabilize a (metastable) trionic Fermi liquid instead of the t-CDW
solution. In our case, this was not an issue due to the fact that the iterative procedure of solution immediately reflects the spontaneous symmetry breaking of the
translational invariance and does not allow for the stabilization of an (unphysical) homogeneous trionic Fermi liquid at half-filling.  

%%%%%%%%%%%%%%%%%%%%%%%%%%%%%%%%%%%%%%%%%%%%%%%%%%%%%%%%%%%%%%%%%%%%%%%%%%%%%%%%%%%%%%%%%%%%%%%%%%%%%%%% 
\begin{figure}
\begin{center}
\vspace{0.05cm}
\includegraphics[scale=0.3]{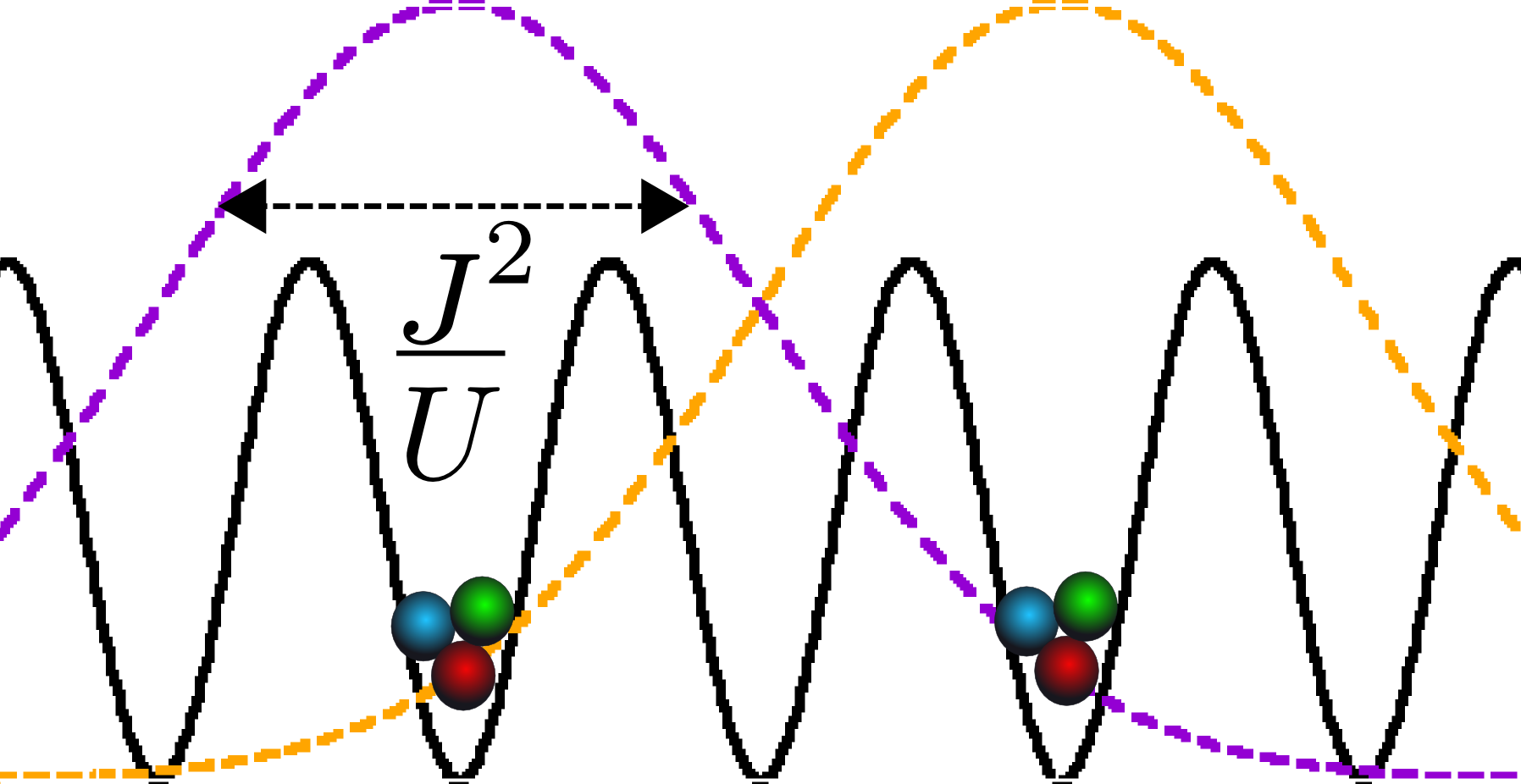}  
\end{center}
\vspace{-0.4cm}
\caption{(Color online) Sketch of the spatial arrangement of trions in the trionic CDW phase.}
\label{cdw_sketch}
\end{figure}
%%%%%%%%%%%%%%%%%%%%%%%%%%%%%%%%%%%%%%%%%%%%%%%%%%%%%%%%%%%%%%%%%%%%%%%%%%%%%%%%%%%%%%%%%%%%%%%%%%%%%%%%

On the other hand, the necessary presence of CDW modulation in the trionic phase at half-filling, at least in the strong-coupling limit, can be easily understood
based on general perturbative arguments. Indeed, as pointed out in Sec. \ref{Methods}, in the strong-coupling trionic phase where $J/|U| \ll 1$, the system
can be described in terms of an effective trionic Hamiltonian (\ref{trion_Hamiltonian}). In this Hamiltonian the effective hopping $J_{eff}$ of the trions is much
smaller than the next-neighbor repulsion $V_{eff}$ between the trions $J_{eff}=\frac{3J^3}{2U^2}\ll V_{eff}=\frac{3J^2}{2|U|}$. Due to the scaling of the hopping
parameter required to obtain a meaningful limit $D\rightarrow\infty$, i.e. $J \to J/\sqrt z$ where $z$ is the lattice connectivity, one finds  $J_{eff} \to 0$ in
this limit, i.e. the trions become immobile  while their next-neighbor interaction term survives. In this limit, the Hamiltonian is equivalent to an 
antiferromagnetic Ising model (spin up corresponds to a trion and spin down corresponds to a trionic-hole). At half-filling, clearly the most energetically favorable
configuration is therefore to arrange the trions in a staggered configuration \cite{Steffen}. Moreover, due to quantum fluctuations, if we decrease the interaction
starting from very large $|U|$, the spread of a single trion (which is proportional to $J^2/U$) increases and it is not a local object any more. In this case the
trionic wave-function extends also to the nearest neighboring sites \cite{Jan}, as sketched in Fig. \ref{cdw_sketch}.  This interpretation is in agreement with the
observed behavior of the CDW order parameter $C$ in Fig. \ref{static}. Indeed, at large $|U|$, $C$ asymptotically rises to the value $C=1/2$, corresponding to the
fully local trions in a staggered CDW configuration. The presence of the CDW also explains the anomalously large value of residual entropy 
per site $s_{res}=k_B\ln 2$ found when imposing a homogeneous trionic phase as in Ref. \cite{inaba,inaba2}. At strong-coupling in finite dimensions, even though the
trions have a finite effective hopping $J_{eff}$, one would still expect that the augmented symmetry at half-filling favors CDW modulations with respect to a trionic
Fermi liquid phase.  In $D=1,2$ it is indeed known \cite{hof1,kantian} that the CDW is actually stable with respect to the SF phase at half-filling for any value of
the interaction, in contrast to the $SU(2)$ case where they are degenerate \cite{toschi}. Our results prove that in higher spatial dimensions this is
not the case and there is a finite range of attraction at weak-coupling, where the c-SF phase is actually stable.

%%%%%%%%%%%%%%%%%%%%%%%%%%%%%%%%%%%%%%%%%%%%%%%%%%%%%%%%%%%%%%%%%%%%%%%%%%%%%%%%%%%%%%%%%%%%%%%%%%%%%%%%
\begin{figure}[htb]
\begin{center}
\subfigure[]{
\begin{minipage}[b]{0.45\textwidth}
\includegraphics[scale=0.275]{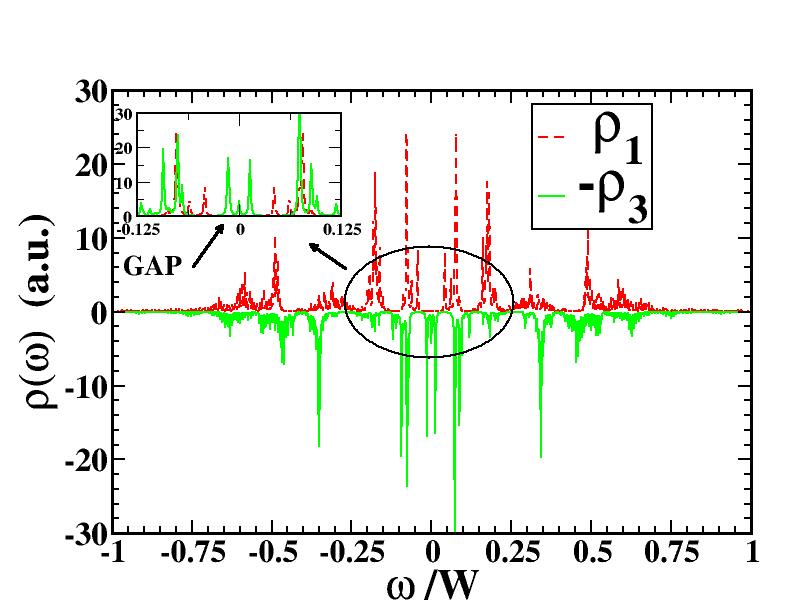} 
\end{minipage}}
\hspace{0.45cm}
\subfigure[]{
\begin{minipage}[b]{0.45\textwidth}
\includegraphics[scale=0.275]{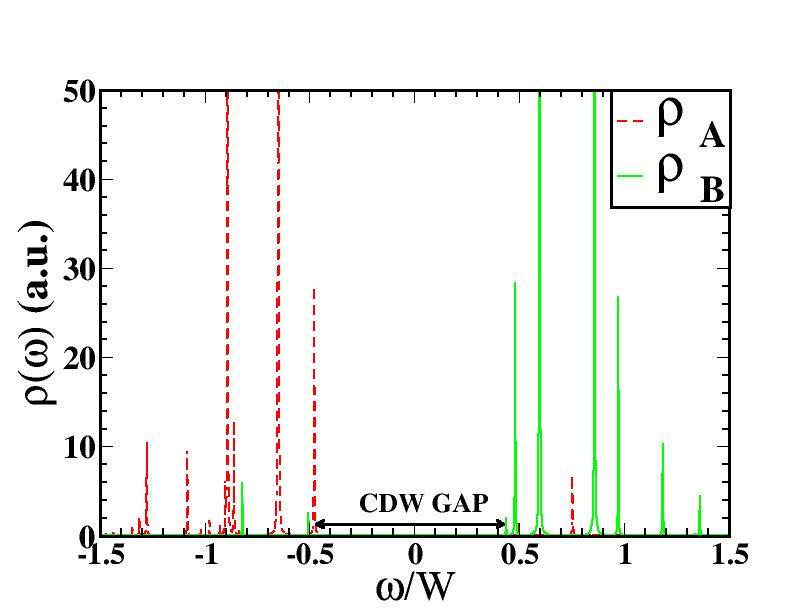}  
\end{minipage}}\\
\end{center}
\vspace{-0.4cm}
\caption{(Color online) Single particle spectral function for the Bethe lattice with $D\rightarrow\infty$ at half-filling and $T=0$ for (a) the c-SF phase at
$|U|/W=0.35$ and (b) the t-CDW phase at $|U|/W=0.75$. In the subfigure (a) we plotted $\rho_1(\omega)$ (red/dashed line) together with $-\rho_3(\omega)$
(green/solid line) to emphasize the different behavior in the paired channel and for the unpaired species. The inset shows the low-energy region and 
the c-SF gap. The subfigure (b) shows the spectral function for sublattices A (red/dashed line) and B (green/solid line) and the gap in the trionic CDW phase.
(Unconstrained, i.e. $V=0$)}
\label{spectral}
\end{figure}
%%%%%%%%%%%%%%%%%%%%%%%%%%%%%%%%%%%%%%%%%%%%%%%%%%%%%%%%%%%%%%%%%%%%%%%%%%%%%%%%%%%%%%%%%%%%%%%%%%%%%%%%

Further confirmation of the physical scenario depicted above is provided by the analysis of the  single-particle spectral function  $\rho_{\sigma}$ in the c-SF and
t-CDW phases shown in Fig.~\ref{spectral}. In the c-SF phase (Fig.~\ref{spectral}(a)), the spectrum shows a gapless branch due to the presence of
the third species which is not involved in the pairing, while the spectral function for species 1 (2 is identical) shows a gap. The situation is totally different in
the t-CDW phase (Fig.~\ref{spectral}(b)), where the spectral functions for the three species are identical but the lattice symmetry is broken into two sublattices.
If we plot the spectral functions for the two sublattices (corresponding to two successive iterations in our DMFT loop)  a CDW gap is visible. We would
like to note that the sharply peaked structure of the spectrum is due to the finite number of orbitals in the ED algorithm. However, the size of the gap should not
be affected significantly by the finite number of orbitals. Interestingly for $|U|=0.75W$ the size of the energy gap $\Delta_{gap}\approx W$ is in very close
agreement with the value obtained within SFA for the same value of the interaction \cite{inaba,inaba2}, indicating that the gap most likely is only weakly affected
by CDW ordering.

%%%%%%%%%%%%%%%%%%%%%%%%%%%%%%%%%%%%%%%%%%%%%%%%%%%%%%%%%%%%%%%%%%%%%%%%%%%%%%%%%%%%%%%%%%%%%%%%%%%%%%%%
\begin{figure}
\begin{center}
\subfigure[]{
\begin{minipage}[b]{0.45\textwidth}
\includegraphics[scale=0.275]{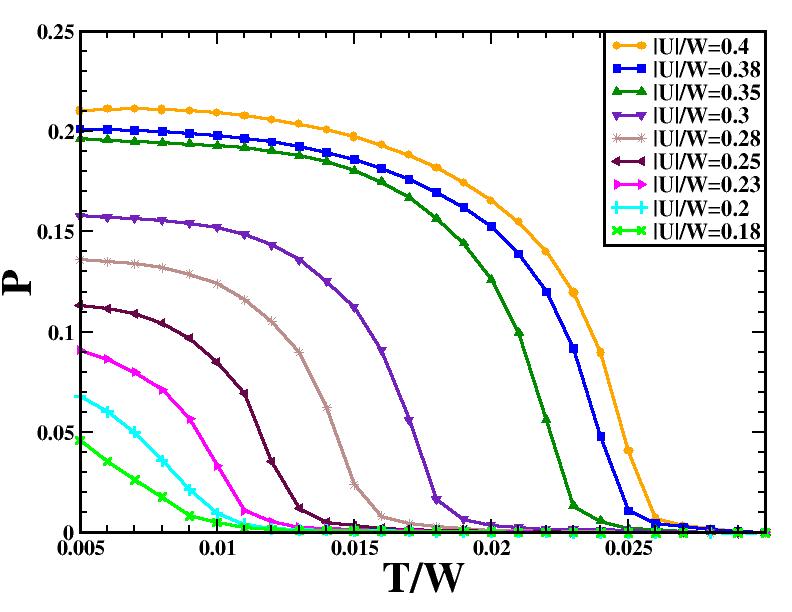}
\end{minipage}}
\hspace{0.45cm}
\subfigure[]{
\begin{minipage}[b]{0.45\textwidth}
\includegraphics[scale=0.275]{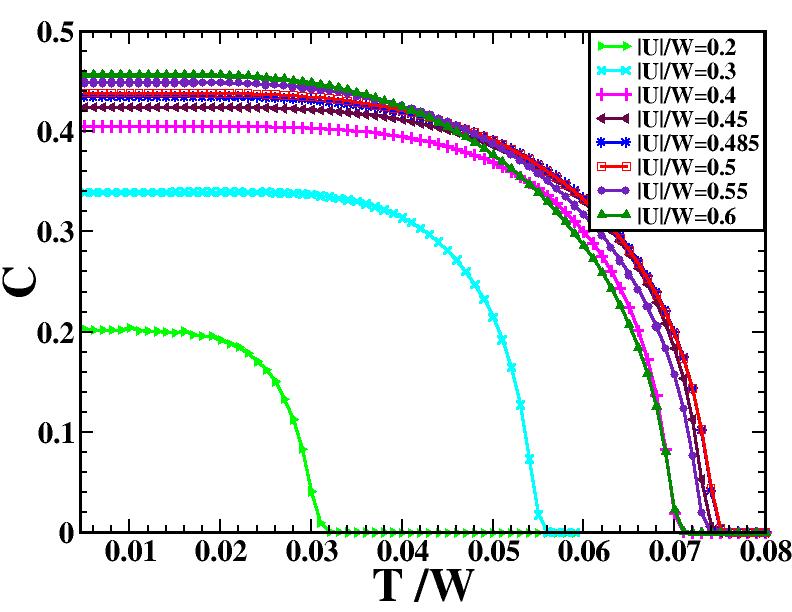}
\end{minipage}}\\
\end{center}
\vspace{-0.4cm}
\caption{(Color online) (a) c-SF order parameter $P$  and (b) CDW amplitude $C$ as a function of temperature $T/W$ on the Bethe lattice  with
$D\rightarrow\infty$ at half-filling. Different lines correspond to different values of the interaction. (Unconstrained, i.e. $V=0$)}
\label{delta_T}
\end{figure}
%\vspace{1cm}
 %%%%%%%%%%%%%%%%%%%%%%%%%%%%%%%%%%%%%%%%%%%%%%%%%%%%%%%%%%%%%%%%%%%%%%%%%%%%%%%%%%%%%%%%%%%%%%%%%%%%%%%%
 
\begin{figure}
\begin{center}
\includegraphics[scale=0.275]{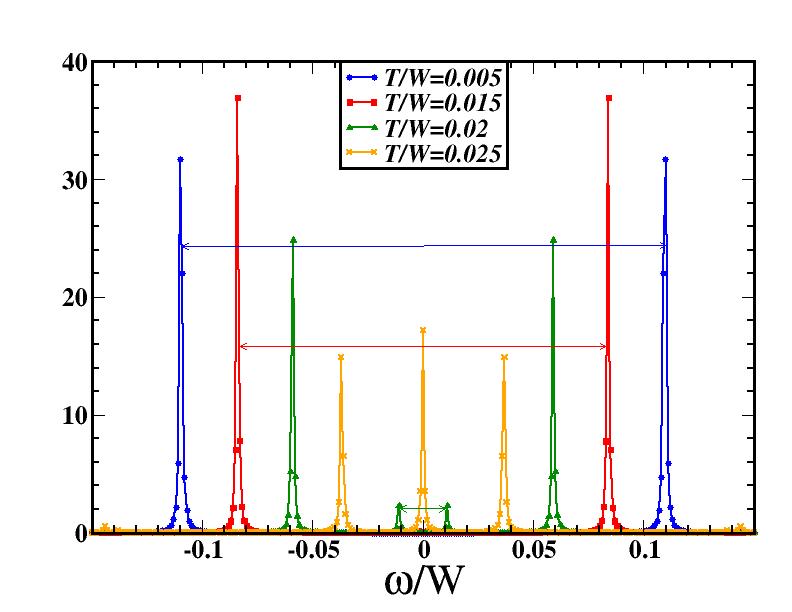}  
\end{center}
\vspace{-0.4cm}
\caption{(Color online) Single particle spectral function on the Bethe lattice with $D\rightarrow\infty$ at half-filling for $|U|/W=0.375$.
Different colors correspond to different values of temperature. (Unconstrained,  i.e. $V=0$) }
\label{spectrum_FT}
\end{figure}

In order to characterize the system at finite temperature, we studied the evolution of the SF order parameter $P$ as a function of temperature in the c-SF
phase for different values of the coupling (Fig. \ref{delta_T}(a)) and analogously for the CDW order parameter $C$ in the t-CDW phase (Fig. \ref{delta_T}(b)). 
The superfluid-to-normal phase transition at $T^{SF}_c(U)$ is also mirrored in the behavior of the spectral function for increasing temperature. The results shown in
Fig. \ref{spectrum_FT} indicate that the superfluid gap in the spectral function closes for $T>T^{SF}_c(U)$, signaling the transition to a normal homogeneus phase
without CDW modulations.

\begin{figure}
\begin{center}
\includegraphics[scale=0.275]{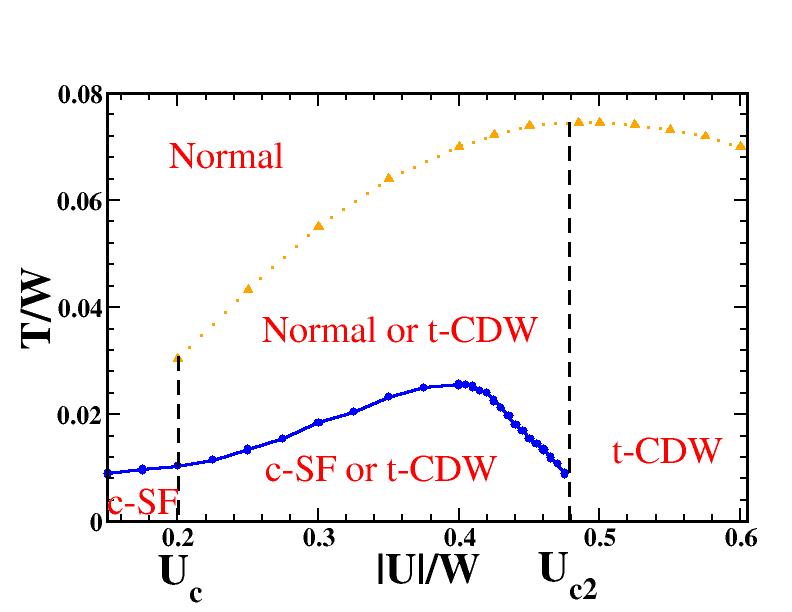}  
\end{center}
\vspace{-0.4cm}
\caption{(Color online) Phase diagram of the unconstrained model ($V=0$) on the Bethe lattice with $D\rightarrow\infty$ at
half-filling as a function of the temperature $T$ and interaction strength $|U|$. The blue solid line $T_c^{SF}$ marks the transition between c-SF to a normal phase,
while the orange dashed line $t_c^{CDW}$ marks the disappearance of CDW modulations in the trionic phase. The dashed vertical lines mark the boundaries of
the coexistence region between the c-SF phase and the t-CDW phase at $T=0$.}
\label{PD}
\end{figure}

At finite temperatures we also found a coexistence region of the trionic CDW wave phase and the color superfluid or normal homogeneous phases in a finite range of
the interaction $U$ ($U_{c1}<|U|<U_{c2}$ at $T=0$). We however leave a thorough investigation of the stability range of the t-CDW phase at finite temperature to
future study, together with its dependence on the distance from the particle-hole symmetric point and on the dimensionality. Due to this coexistence region, we define
the two critical temperatures $T^{SF}_c(U)$ and $T_c^{CDW}(U)$ plotted in the phase diagram in Fig. \ref{PD}, where $P(T)_{|U}$ and $C(T)_{|U}$ vanish respectively
above the c-SF phase and t-CDW phase.  In agreement with the results obtained within SFA \cite{inaba,inaba2}, we also found that the critical temperature $T^{SF}_c(U)$ has
a maximum at $T^{SF}_c/W \approx 0.025$ for $|U|/W =0.4$. This is also in qualitative agreement with the $SU(2)$ case \cite{toschi}, where the critical temperature
has a maximum at intermediate couplings. Due to the presence of the CDW modulations in the trionic phase which are ignored in Ref. 
\cite{inaba,inaba2}, we found also a
second critical temperature $T_c^{CDW}$ where charge density wave modulations in the trionic phase disappear.

%%%%%%%%%%%%%%%%%%%%%%%%%%%%%%%%%%%%%%%%%%%%%%%%%%%%%%%%%%%%%%%%%%%%%%%%%%%%%%%%%%%%%%%%%%%%%%%%%%%%%%%%
%%%%%%%%%%%%%%%%%%%%%%%%%%%%%%%%%%%%%%%%%%%%%%%%%%%%%%%%%%%%%%%%%%%%%%%%%%%%%%%%%%%%%%%%%%%%%%%%%%%%%%%%

%%%%%%%%%%%%%%%%%%%%%%%%%%%%%%%%%%%%%%%%%%%%%%%%%%%%%%%%%%%%%%%%%%%%%%%%%%%%%%%%%%%%%%%%%%%%%%%%%%%%%%%%
%%%%%%%%%%%%%%%%%%%%%%%%%%%%%%%%%%%%%%%%%%%%%%%%%%%%%%%%%%%%%%%%%%%%%%%%%%%%%%%%%%%%%%%%%%%%%%%%%%%%%%%%

\subsection{Incommensurate density}

In this section we consider the system for densities far from the particle-hole symmetric point. 
Specifically we investigate, using VMC and DMFT respectively, the
implementation of the model (\ref{hamiltonian}) on a simple-square (cubic) lattice in 2D (3D) with tight-binding dispersion, i.e. $
 \epsilon_{\bf{k}}=-2J\sum_{i=x,y(,z)} \cos(k_i a)$, where $a$ is the lattice spacing. In particular, we will find that away from
the particle-hole symmetric point in the c-SF phase, the superfluidity always triggers a density imbalance, i.e. a magnetization.

In order to address this feature quantitatively, we studied the system by adjusting the chemical potential $\mu$
in order to fix the \emph{total} density $n=\sum_\sigma n_\sigma$, allowing the system to adjust spontaneously 
the densities in each channel. Due to the \emph{spontaneous} symmetry breaking of the $SU(3)$
symmetry of the Hamiltonian in the color superfluid phase, it is indeed possible that, for a given chemical
potential $\mu_1=\mu_2=\mu_3=\mu$, the particle densities for different species may differ. If such a situation
occurs, the systems shows a finite onsite magnetization $m$.  As a more technical remark, we add that the choice of pairing channel,
as explained in Sec. \ref{sec:thset}, is done without loss of generality: A specific choice will therefore determine in
which channel a potential magnetization takes place, but not influence its overall occurrence. Here, since we fix the pairing
to occur between species 1 and 2, we found a nonzero value of the magnetization parameter $m = n_{12} - n_3$,
where $n_{12} = n_1 = n_2$. Therefore the paired channel turns out  (spontaneously) to be fully balanced, 
while there is in general a finite density imbalance between particles in the paired channel with respect
to the unpaired fermions. 

The implications of the results presented in this subsection and in Sec. \ref{constrained} 
for cold atom experiments, where the total number of particles of each species 
$N_\sigma =\sum_{i} n_{i,\sigma} $ is fixed, will be discussed in Sec. \ref{Phase_Separation}.  
Combining the grand canonical DMFT results with energetic arguments 
based on canonical VMC calculations, we show that the system is 
generally unstable towards domain formation.

%%%%%%%%%%%%%%%%%%%%%%%%%%%%%%%%%%%%%%%%%%%%%%%%%%%%%%%%%%%%%%%%%%%%%%%%%%%%%%%%%%%%%%%%%%%%%%%%%%%%%%%%
\begin{figure}
\begin{center}
\includegraphics[scale=0.275]{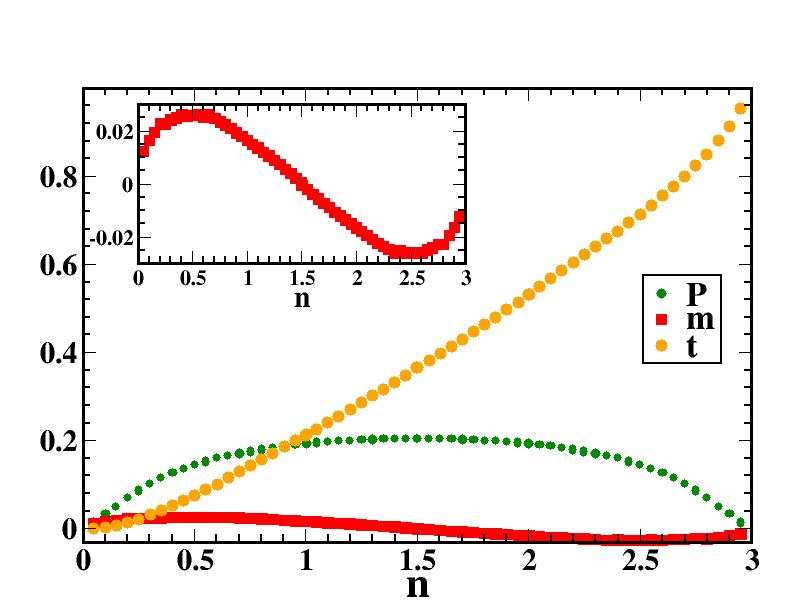}  
\end{center}
\vspace{-0.4cm}
\caption{(Color online) c-SF order parameter $P$ (green circles), magnetization $m$ (red squares) and average triple occupancy 
$t=\langle n_1n_2n_3\rangle$  (violet diamonds) plotted as a function of the total density $n$  per lattice site for $|U/W|=0.3125$ and $T=0$ on 
the cubic lattice in $D=3$. The inset shows the behavior of the magnetization in detail. (Unconstrained, i.e. $V=0$)}
\label{U3.75_V0}
\end{figure}
%%%%%%%%%%%%%%%%%%%%%%%%%%%%%%%%%%%%%%%%%%%%%%%%%%%%%%%%%%%%%%%%%%%%%%%%%%%%%%%%%%%%%%%%%%%%%%%%%%%%%%%%

We first consider in  Fig. \ref{U3.75_V0} how the ground state properties of the $3D$ system evolve by fixing the coupling at $|U|/W = 0.3125$, where the system is
always found to be in the c-SF phase for any density. We consider only densities ranging from $n=0$ to half-filling $n=1.5$. The results above half-filling can be
easily obtained exploiting a particle-hole transformation. In particular one easily obtains
\begin{eqnarray}
P(n)=P(3-n)\ \ \mbox{and}\ \ m(n)=-m(3-n),\\
t(n)=-t(3-n)+n-2+d(3-n) ,
\end{eqnarray}
%\ \ \mbox{and}\ \ d(n)=d(3-n)+2n-3
where $t$ and $d$ are the average triple and double occupancies.  The superfluid order parameter $P$ increases (decreases) with the density for $n < 1.5~(n>1.5)$
and is maximal at half-filling. The average triple occupancy is instead a monotonic function of the density. Below half-filling, the magnetization $m$ first
grows with increasing density, then reaches a maximum and eventually decreases and vanishes at half-filling in agreement with the findings in the previous
subsection. This means that in the c-SF phase for a fixed value of the chemical potential $\mu$ the system favors putting more particles into the paired channel
than into the unpaired component. For $n > 1.5$ the effect is the opposite and $m<0$. This behavior can be understood by considering that
the equilibrium value of the magnetization results from a competition between the condensation energy gain in the paired channel on one side and the potential
energy gain on the other side. Indeed the condensation energy found as a function of the density of pairs has a maximum at half-filling. For example in the
weak-coupling BCS regime $E_{cond}$ is proportional to $P^2 $\cite{baym}. Therefore the condensation energy gain will increase by choosing the number of particles in
the paired channel as close as possible to half-filling. On the other hand, for a fixed total density $n$, this would reduce or increase the
unpaired fermions and consequently the potential energy gain, which is maximal for a non-magnetized system since $U$ is negative. The competition between these
opposite trends eventually determines the value of the magnetization in equilibrium, which is finite and rather small at this value of the coupling (see inset in
Fig. \ref{U3.75_V0}). At half-filling no condensation energy gain can be achieved by creating a density imbalance between the superfluid pairs and the unpaired
fermions since the condensation energy is already maximal. Therefore the spontaneous symmetry breaking in the color superfluid phase does not 
result necessarily in a density imbalance, which is however triggered by a condensation energy gain for every density deviation from the 
particle-hole symmetric point.

%%%%%%%%%%%%%%%%%%%%%%%%%%%%%%%%%%%%%%%%%%%%%%%%%%%%%%%%%%%%%%%%%%%%%%%%%%%%%%%%%%%%%%%%%%%%%%%%%%%%%%%%
\begin{figure}
\begin{center}
\includegraphics[scale=0.275]{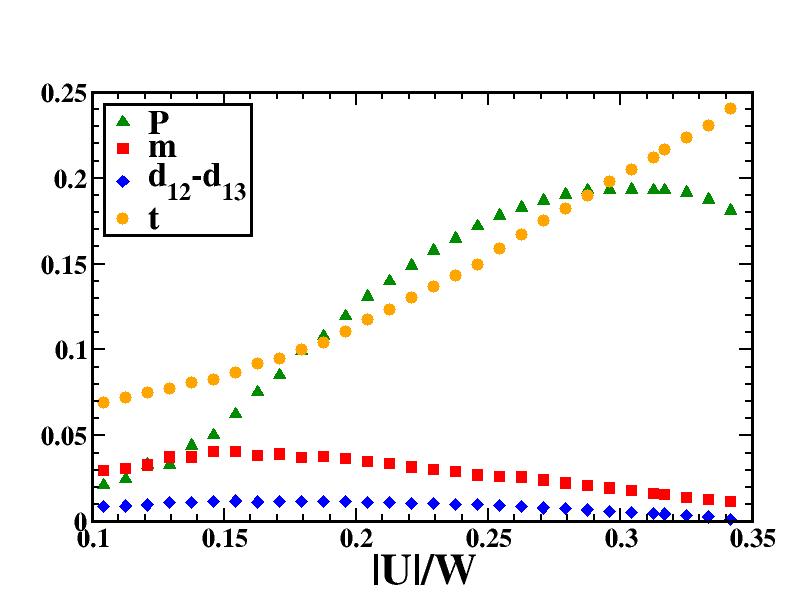}  
\end{center}
\vspace{-0.4cm}
\caption{(Color online) c-SF order parameter $P$ (green triangles), magnetization $m$ (red squares), average triple occupancy 
$t=\langle n_1n_2n_3\rangle$ (orange circles) and difference between double occupancies in different channels $d_{12}-d_{13}$ (blue 
diamonds) in the c-SF phase, plotted as a function of the interaction $|U/W|$ for $n=1$ and $T=0$ for the cubic lattice in $D=3$.
(Unconstrained,  i.e. $V=0$)}
\label{V0_n1}
\end{figure}
%%%%%%%%%%%%%%%%%%%%%%%%%%%%%%%%%%%%%%%%%%%%%%%%%%%%%%%%%%%%%%%%%%%%%%%%%%%%%%%%%%%%%%%%%%%%%%%%%%%%%%%%

We now consider the same system for fixed total density $n=1$ and study the ground state properties as a function
of the interaction strength $|U|$ (see Fig.\ref{V0_n1}). For weak interactions the system is in a c-SF phase. Upon increasing $|U|$, the order parameter $P$ first increases and then shows the dome shape at intermediate
couplings which we already observed for the half-filled case. Away from the half-filling, the value where $P$ reaches its maximum is shifted to lower values of the
interaction strength. The triple occupancy $t$, on the other hand monotonically increases with $|U|$. Interestingly the magnetization $m(U)$ has a non-monotonic
behavior. At weak-coupling, magnetization $m(U)$ grows with increase of the interaction strength. For increasing coupling, $m$ has a maximum and then decreases for
larger $|U|$, indicating a non-trivial evolution due to competition between the condensation energy and the potential energies for increasing attraction. The
spontaneous breaking of the $SU(3)$-symmetry is also well visible in the behavior of the double occupancies. Indeed in the c-SF for $n<1.5$ we find $d_{12}>
d_{13}=d_{23}$. The difference $d_{12}-d_{23}$ is however non-monotonic in the coupling and seems to vanish at $|U|/W\approx 0.35$. Our interpretation
is that beyond this point the $SU(3)$-symmetry is restored and the system undergoes a transition to a Fermi liquid trionic phase. Indeed for $|U|/W > 0.35$ we did
not find any converged solution within our DMFT approach, neither for a homogeneous nor for a staggered criterion of convergence. This result is compatible with
the presence of a macroscopically large number of degenerate trionic configurations away from the half-filling. A finite kinetic energy for the trions would remove
this degeneracy, leading to a trionic Fermi liquid ground state. This contribution is however beyond the DMFT description of the trionic phase where trions are
immobile objects. We can address the existence of a Fermi liquid trionic phase at strong-coupling using the VMC approach in $2D$, which we will
discuss in the following.

%%%%%%%%%%%%%%%%%%%%%%%%%%%%%%%%%%%%%%%%%%%%%%%%%%%%%%%%%%%%%%%%%%%%%%%%%%%%%%%%%%%%%%%%%%%%%%%%%%%%%%%%
\begin{figure}
\begin{center}
\includegraphics[scale=0.275]{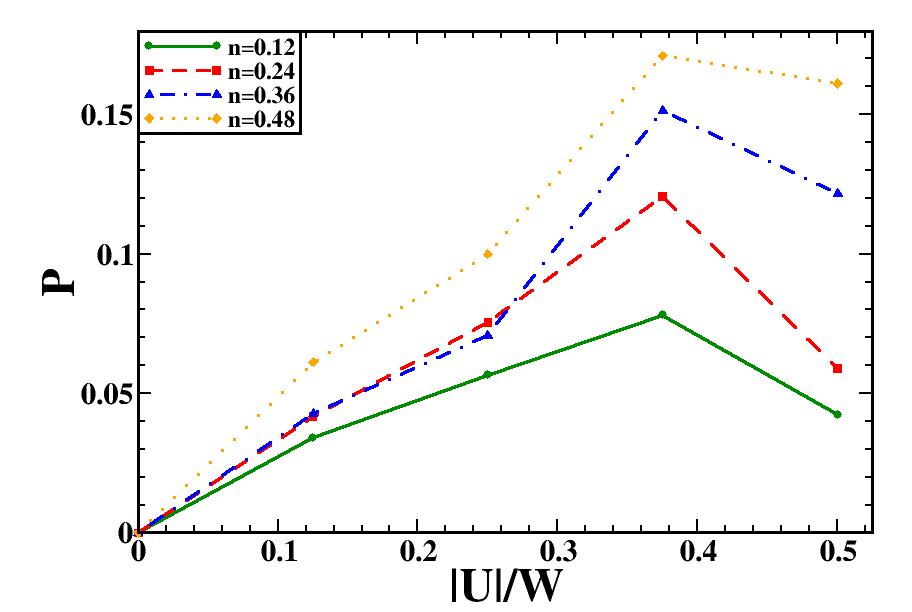}  
\end{center}
\vspace{-0.4cm}
\caption{(Color online) Superfluid order parameter on the 2D square lattice for different total filling as a function of the interaction strength. We
 neglect spontaneous magnetization in the system. (Unconstrained, i.e. $V=0$)}
\label{Unconstrained_SC_and_Condensation_energy}
\end{figure}
%%%%%%%%%%%%%%%%%%%%%%%%%%%%%%%%%%%%%%%%%%%%%%%%%%%%%%%%%%%%%%%%%%%%%%%%%%%%%%%%%%%%%%%%%%%%%%%%%%%%%%%%
%%%%%%%%%%%%%%%%%%%%%%%%%%%%%%%%%%%%%%%%%%%%%%%%%%%%%%%%%%%%%%%%%%%%%%%%%%%%%%%%%%%%%%%%%%%%%%%%%%%%%%%%
\begin{figure}
\begin{center}
\includegraphics[scale=0.275]{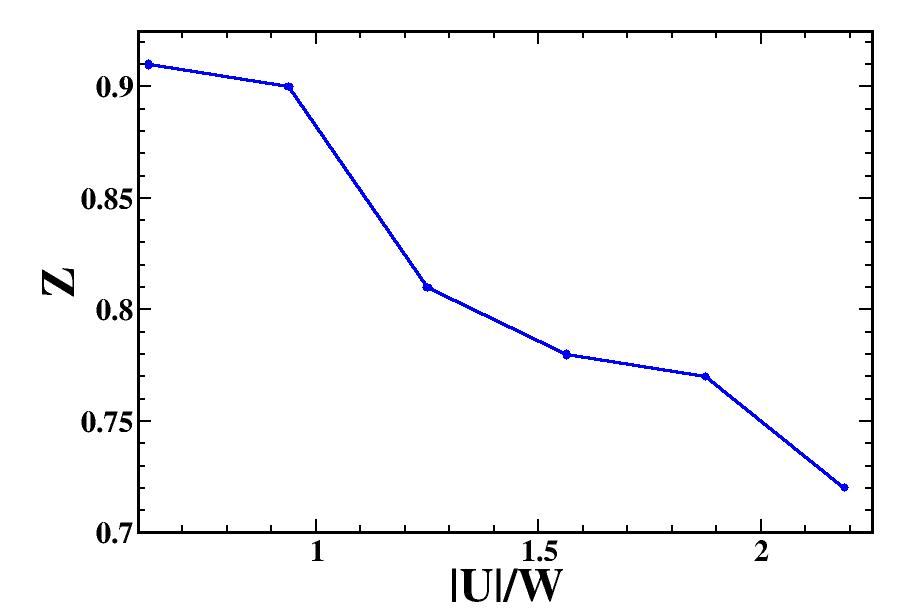}  
\end{center}
\vspace{-0.4cm}
\caption{(Color online) The quasi-particle weight $Z$ averaged over the Fermi surface as a function of the interaction strength $|U|$.
 (Unconstrained, i.e. $V=0$)}
\label{quasi-particle_weight}
\end{figure}
%%%%%%%%%%%%%%%%%%%%%%%%%%%%%%%%%%%%%%%%%%%%%%%%%%%%%%%%%%%%%%%%%%%%%%%%%%%%%%%%%%%%%%%%%%%%%%%%%%%%%%%%
As already mentioned in Sec. \ref{Methods}, we use different trial wavefunctions to study the behavior of the system in the weak- ($|U| \leq W/2$) and the
strong-coupling ($|U| > W/2$) regimes. At weak-coupling the magnetization is expected to be very small and we can consider the results for the unpolarized system
with $n_1=n_2=n_3$ to be a good approximation of the real system which is in general polarized. We found indeed that for $|U| \leq W/2$ the system is in the c-SF
phase with a finite order parameter $P$. As shown in  Fig. \ref{Unconstrained_SC_and_Condensation_energy}, we obtain that $P(U)$ has a similar dome shape as in the
3D case. Unfortunately, we cannot  directly address the trionic transition within this approach since it is expected to take place at intermediate coupling where
both ansatz wave functions are inaccurate. We can however consider the system in the strong-coupling limit by using the effective trionic Hamiltonian of
Eq.~\ref{trion_Hamiltonian}. In this way we can study the Fermi liquid trionic phase 

which we characterize by evaluating the quasiparticle weight, averaged over the Fermi surface 
%\begin{equation}
%\label{quasi-particle-weight} 
%Z=\frac{\sum_{\bf{k}}Z_{\bf{k}}\tilde\delta(\varepsilon_{\bf k}-E_F)}{\sum_{\bf{k}} \tilde\delta(\varepsilon_{\bf k}-E_F)} 
%\quad\quad {\rm   where\quad  
%\tilde\delta(\varepsilon_{\bf k}-E_F)=\left\{
%\begin{array}{cc}
%1&\varepsilon_{\bf k}=E_F\\
%0&{\rm otherwise} 
%\end{array} \right. 
%\, .
%\end{equation}

\begin{equation}
\label{quasi-particle-weight} 
Z=\frac{\sum_{\bf{k}}Z_{\bf{k}} \delta_{\varepsilon_{\bf k},E_F}}{\sum_{\bf{k}} \delta_{\varepsilon_{\bf k},E_F}} \, .
\end{equation} 
Here $Z_{\bf k}$ is extracted from the jump in the momentum distribution at the Fermi surface, which we approximate as
\begin{equation}
\label{z_k_quasi-particle} 
Z_{\bf k}=n_{\bf k} - \frac{1}{2}\left(n_{{\bf k}+\Delta k_x}+n_{{\bf k}+\Delta k_y}\right) \, ,
\end{equation}
where $\Delta{\bf k}_{x}$  ($\Delta{\bf k}_{y}$) is the translational vector along the $x$ ($y$) direction in the reciprocal lattice. In
Fig.~\ref{quasi-particle_weight} we plot $Z$ as a function of interaction strength $|U|/W$.

By combining DMFT and VMC results we therefore have strong evidence of the system undergoing a phase transition from a magnetized color-superfluid to a trionic Fermi
liquid phase at strong-coupling, when the density is far enough from the particle-hole symmetric point. 
 
%%%%%%%%%%%%%%%%%%%%%%%%%%%%%%%%%%%%%%%%%%%%%%%%%%%%%%%%%%%%%%%%%%%%%%%%%%%%%%%%%%%%%%%%%%%%%%%%%%%%%%%%
\begin{figure}
\begin{center}
\includegraphics[scale=0.275]{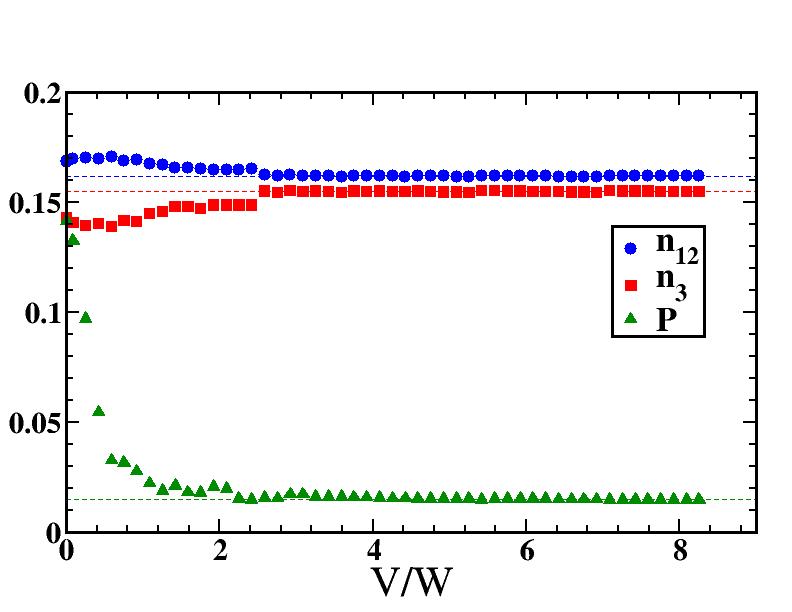}  
\end{center}
\vspace{-0.4cm}
\caption{(Color online) Number of particles in the paired channels $n_{12}=n_1=n_2$ (blue circles) and the unpaired channel $n_3$ (red squares) and superfluid order
parameter $P$ as a function of the 3-body repulsion $V$ for $|U|/W=0.312$ and total density $n=0.48$ for the cubic lattice in
$D=3$ at zero
temperature. Dashed lines correspond to the asymptotic values. }
\label{U3.75_dens0.48_Vvar}
\end{figure}
%%%%%%%%%%%%%%%%%%%%%%%%%%%%%%%%%%%%%%%%%%%%%%%%%%%%%%%%%%%%%%%%%%%%%%%%%%%%%%%%%%%%%%%%%%%%%%%%%%%%%%%%

\section{Results: Constrained System ($V=\infty$)}\label{constrained}

As referred to in the introduction, actual laboratory implementations of the model under investigation
 using ultracold gases are often affected with three-body losses,
which are not Pauli suppressed as in the $SU(2)$ case. As discussed in Ref. \cite{selim}, the three-body loss rate $\gamma_3$ shows a strong dependence on the
applied magnetic field. Therefore the results presented in the previous section essentially apply to the case of cold gases only whenever three-body losses are
negligible, i.e. $\gamma_3\ll J,U$. In the general case, in order to model the system in presence of three-body losses, one needs a non-equilibrium formulation where
the number of particles is not conserved. However, as shown in Ref. \cite{daley}, in the regime of strong losses $\gamma_3 \gg J,U$, the probability of having triply
occupied sites vanishes and the system can still be described using a Hamiltonian formulation with a dynamically-generated three-body constraint. To take it into
account in our DMFT formalism, we introduce a three-body repulsion with $V=\infty$. Within VMC we directly project triply occupied sites out of the Hilbert space.
We stress that finite values of $V$ do not correspond to real systems with moderately large $\gamma_3$ since then real losses occur and a purely Hamiltonian
description does not apply any more; only the limits $\gamma_3\ll J,U$ and $\gamma_3\gg J,U$ lend themselves to an effective Hamiltonian formulation. 

\subsection{Ground State Properties}

%%%%%%%%%%%%%%%%%%%%%%%%%%%%%%%%%%%%%%%%%%%%%%%%%%%%%%%%%%%%%%%%%%%%%%%%%%%%%%%%%%%%%%%%%%%%%%%%%%%%%%%%
\begin{figure}
\begin{center}
\includegraphics[scale=0.275]{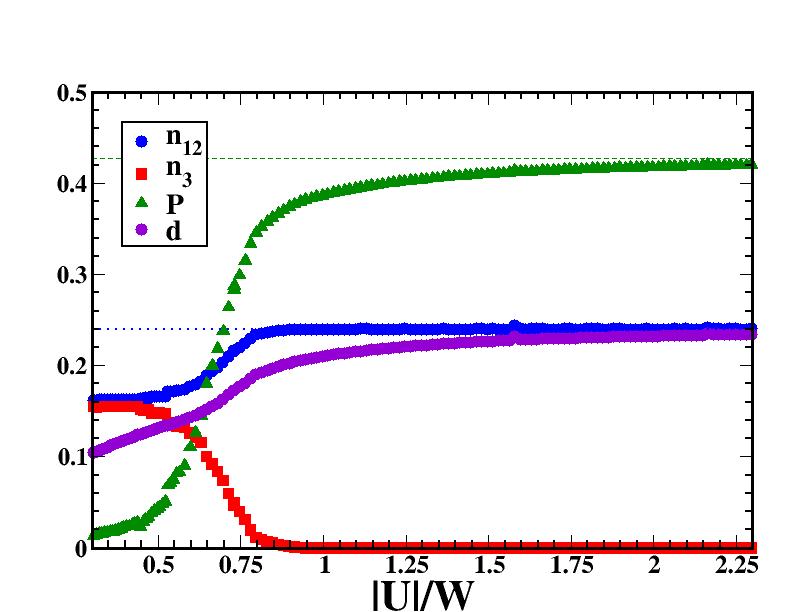}  
\end{center}
\vspace{-0.4cm}
\caption{(Color online) Number of particles for the paired channels $n_{12}=n_1=n_2$ (blue/dark circles)  and the unpaired channel $n_3$ (red squares), c-SF order
parameter $P$ (green triangles) and total double occupancy $d=d_{12}+d_{13}+d_{23}$ (violet/light circles) calculated within DMFT
as a function of the interaction strength $|U|/W$ for $T=0$, $V\approx 80 W$ and $n=0.48$ (cubic lattice). The dashed green line corresponds to the asymptotic value of the superfluid order parameter in the atomic
limit $P_{\infty}$, while the dotted blue line corresponds to the asymptotic value of the particle density in the paired channel, which is also equal to the
asymptotic value of the total 
double occupancy. (Constrained case, $V\simeq 80W$)}
\label{dens0.48_V100}
\end{figure}
 %%%%%%%%%%%%%%%%%%%%%%%%%%%%%%%%%%%%%%%%%%%%%%%%%%%%%%%%%%%%%%%%%%%%%%%%%%%%%%%%%%%%%%%%%%%%%%%%%%%%%%%%
In order to address how the system approaches the constrained regime with increasing $V$, we first used DMFT to study the ground-state properties of the model in
$3D$ as a function of  the three-body interaction $V$ for a fixed value of the total density $n=0.48$ and the two-body attraction $|U|/W=0.3125$. We found that the
average number of triply occupied sites $t=\langle n_1n_2 n_3\rangle$ (not shown) vanishes very fast with increasing $V$. The SF order parameter $P$ and the
densities in the paired and unpaired channels approach their asymptotic values already for $V\approx 3W$ or $V \approx 10 |U|$, as shown in 
Fig.~\ref{U3.75_dens0.48_Vvar} Therefore, we assume that we can safely consider the system to be in the constrained regime whenever $V$ is chosen to be much larger
than this value. 

%%%%%%%%%%%%%%%%%%%%%%%%%%%%%%%%%%%%%%%%%%%%%%%%%%%%%%%%%%%%%%%%%%%%%%%%%%%%%%%%%%%%%%%%%%%%%%%%%%%%%%%%
\begin{figure}[hbpt]
\begin{center}
\includegraphics[scale=0.275]{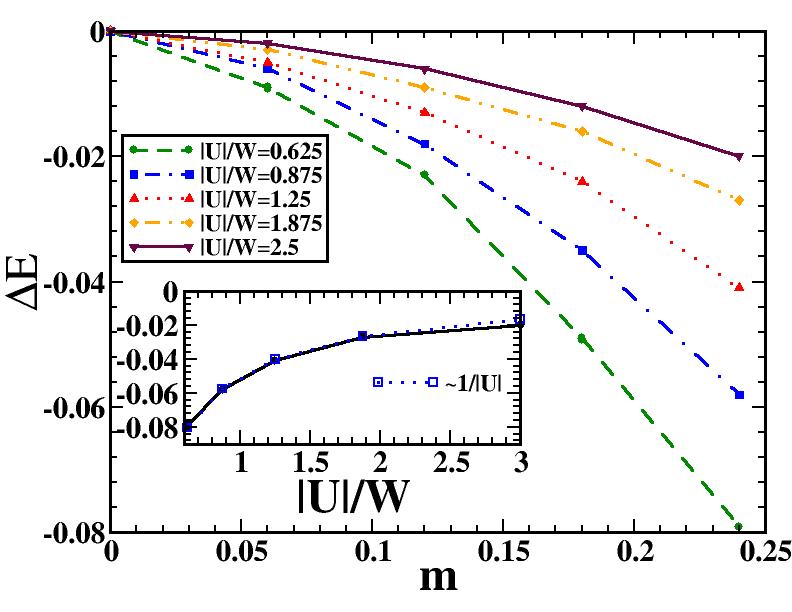}
\end{center}
\caption{Effect of the magnetization for total density $n=0.48$ on the 2D square lattice with 50 lattice sites. In particular we plot $\Delta E(m)=E(m)-E(0)$
as a function of magnetization $m=n_{12}-n_3$ ($n_{12}=n_1 = n_2)$ for different values of the interaction strength $U$. 
Calculations are performed using the VMC method with a strong coupling ansatz. In the inset we plot $\Delta E$ as a function 
of the interaction strength $|U|$ for the fully magnetized c-SF phase. (Constrained case)}
\label{Energy_difference}
\end{figure}
%%%%%%%%%%%%%%%%%%%%%%%%%%%%%%%%%%%%%%%%%%%%%%%%%%%%%%%%%%%%%%%%%%%%%%%%%%%%%%%%%%%%%%%%%%%%%%%%%%%%%%%%

Both the densities $n_\sigma$ and the superfluid order parameter $P$ are strongly affected by the three-body interaction 
(see Fig. \ref{U3.75_dens0.48_Vvar}). For this value of the interaction, $P$ and $m$ are strongly suppressed by the three-body repulsion, even though both eventually saturate to a finite value for large
enough $V$. However, as shown below, this suppression of the magnetization and SF properties is specific to the weak-coupling regime and for larger values of $|U|$
both the SF order parameter $P$ and the magnetization $m$ are instead strongly enhanced in the presence of large $V$.

We now investigate the constrained case (setting $V=1000J\approx 80 W$ within the DMFT approach)  %but results are indistinguishable from $V=100t\approx 8W$) 
where the \emph{total} density is fixed as above to $n=0.48$. Large values of the density imply 
an increase of the probability of real losses over a finite interval of time.
Therefore we restrict ourselves to a relatively
low density which is meant to be representative of a possible experimental setup.  

We study the evolution of the ground state of the system in $2D$ and $3D$ as a function of the two-body
interaction strength $U$. DMFT results in Fig. \ref{dens0.48_V100} show that in the
three-dimensional system the trionic phase at strong coupling is completely suppressed by the three-body constraint
and the ground state is found to be always a color superfluid for any value of the attraction.
This remaining c-SF phase shows however a very peculiar behavior of
the magnetization $m$ as a function of the attraction $U$.
Indeed the magnetization $m = n_{12} - n_3$ ($n_{12}=n_1=n_2$) steadily increases for increasing interaction 
and $n_3 \approx 0$ ($m \approx n_{12} \approx n/2$) already for $U \approx 12J=W$. 

Our explanation is that the three-body constraint strongly affects the energetic balance within the c-SF phase. Indeed, in the 
absence of $V$ the magnetization was shown to be non-monotonic and to vanish in the $SU(3)$-symmetric trionic phase at 
strong-coupling. Now instead in the same limit the fully polarized c-SF system has a smaller ground state energy for fixed total
density $n$. This result is fully confirmed by the VMC data for the 2D square lattice. As shown in the
next section, combining these results essentially implies that a globally homogeneous phase with $m=0$
is unstable in the thermodynamic limit with respect to domain formation whenever the global particle \emph{number} 
in each species $N_\sigma = \sum_i n_{i,\sigma}$ is conserved. By using the canonical
ensemble approach of VMC, we can indeed address also metastable phases and study the effect on the energy of a finite magnetization for fixed total density $n=0.48$.
In particular we study the energy difference between the magnetized system and the unpolarized one with the same $n$, i.e. $\Delta E(m)=E(m)-E(0)$.  Results shown in
Fig. \ref{Energy_difference} indicate that at strong-coupling the energy decreases for increasing magnetization and the minimum in the ground state energy
corresponds to the fully polarized system. In the inset of Fig.~\ref{Energy_difference}, we show $\Delta E$ as a function of the interaction strength for the
fully polarized c-SF at strong-coupling, which decreases as  $\Delta E \sim 1/|U|$. We also investigated the system in the weak-coupling regime, where
our calculation shows that  $\Delta E(m)$ has a minimum for very small values of the magnetization (not shown). This indicates that also in $2D$ the c-SF ground
state at weak-coupling is partially magnetized, in complete agreement with the three-dimensional results.    

%%%%%%%%%%%%%%%%%%%%%%%%%%%%%%%%%%%%%%%%%%%%%%%%%%%%%%%%%%%%%%%%%%%%%%%%%%%%%%%%%%%%%%%%%%%%%%%%%%%%%%%%
\begin{figure}[hbpt]
\begin{center}
\subfigure[]{
\begin{minipage}[b]{0.35\textwidth}
\hspace{-3.5cm}
\includegraphics[scale=0.275]{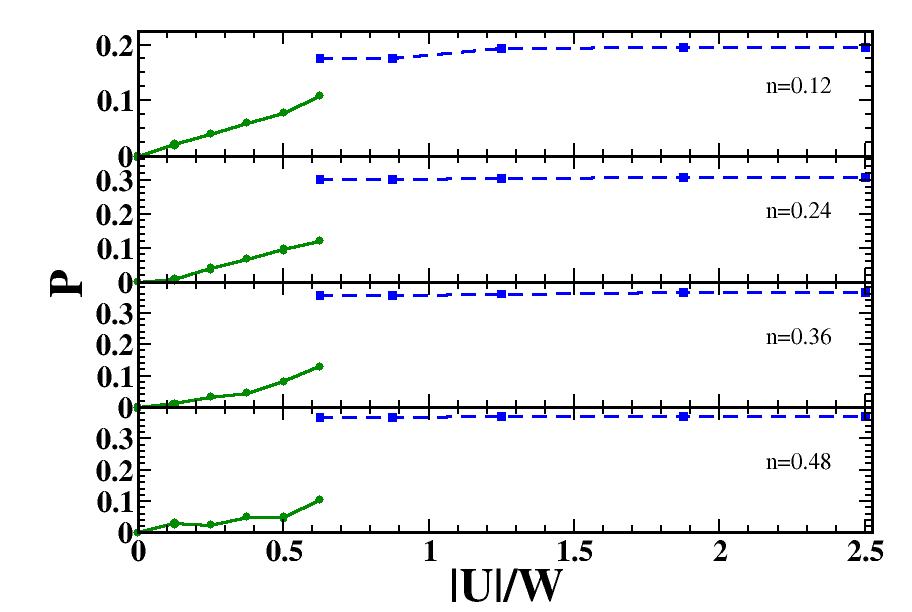}
\end{minipage}}
\subfigure[]{
\begin{minipage}[b]{0.35\textwidth}
\hspace{-0.5cm}
\includegraphics[scale=0.275]{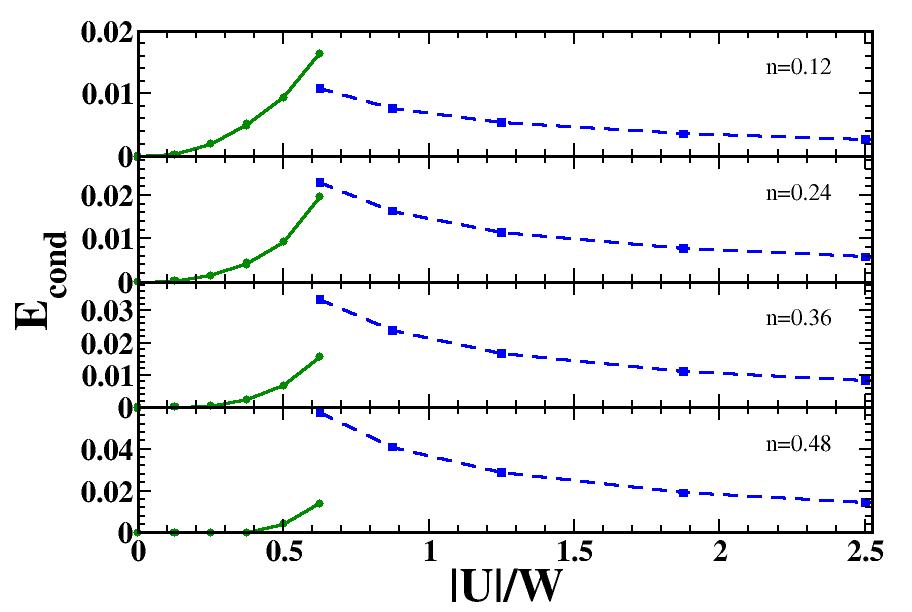}
\end{minipage}}\\
\end{center}
\caption{(a) The superfluid order parameter and (b) condensation energy for 2D square lattice for different total fillings as a function of
the interaction strength. For weak coupling we approximate that the system is not magnetized (green lines and circles), while for strong
coupling we assume that the system
is fully polarized, i.e. contains only pairs (dashed blue line and squares). The dotted line corresponds to the superfluid order parameter in the atomic limit $P_\infty$
(Constrained case)} 
\label{SC_and_Condensation_energy}
\end{figure}
%%%%%%%%%%%%%%%%%%%%%%%%%%%%%%%%%%%%%%%%%%%%%%%%%%%%%%%%%%%%%%%%%%%%%%%%%%%%%%%%%%%%%%%%%%%%%%%%%%%%%%%%

Within DMFT the order parameter $P$ in the c-SF ground state shown in Fig. \ref{dens0.48_V100} is also increasing with $|U|$ and saturates at strong coupling to a
finite value, which we found to be in agreement with the asymptotic value in the atomic limit for the $SU(2)$ symmetric case \cite{toschi}
\begin{equation}
P_{\infty}=\lim_{U/W \to \infty} P(U)=\frac{1}{2}\sqrt{n(2-n)} \,.
\end{equation}

The total number of double occupancies $d$ is also an increasing function of $|U|$ and saturates for very large $|U|$ to the value $n_{12}=n/2$ as in the strong
coupling limit for the $SU(2)$ symmetric system. This means that in the ground state the strong coupling limit of the $SU(3)$ model is
indistinguishable from the $SU(2)$ case for the same total density $n$ and two-body interaction $U$. As we will show in the
next subsection, this is not any more true if we consider instead finite temperatures. 

Similar considerations on the superfluid properties in the ground state apply to the two-dimensional case studied within the VMC
technique. As the magnetization in the weak-coupling regime is very small, we approximated it to zero and consider an unpolarized
system within the weak-coupling ansatz, while at strong-coupling we directly consider the system as fully polarized, i.e. containing
only pairs. As visible in Fig. \ref{SC_and_Condensation_energy}, $P$ shows a similar behavior to
the 3D case. Indeed at weak-coupling both, DMFT and VMC, show a BCS exponential behavior in the coupling,
while at strong-coupling $P$ converges to a constant.

Within VMC we also studied the condensation energy as explained in Sec. \ref{Methods}. Fig. \ref{SC_and_Condensation_energy}b shows
that the condensation energy first increases with the interaction strength $U$ as expected in BCS theory, while it decreases as $1/U$
at strong-coupling as expected in the BEC limit for the $SU(2)$ case \cite{toschi}. Despite the fact that we cannot reliably address
the intermediate region, there are also indications that the condensation energy has a
maximum in this region.

\subsection{Finite temperatures}

%%%%%%%%%%%%%%%%%%%%%%%%%%%%%%%%%%%%%%%%%%%%%%%%%%%%%%%%%%%%%%%%%%%%%%%%%%%%%%%%%%%%%%%%%%%%%%%%%%%%%%%%
\begin{figure}[hbpt]
\begin{center}
\subfigure[]{
\begin{minipage}[b]{0.45\textwidth}
\includegraphics[scale=0.275]{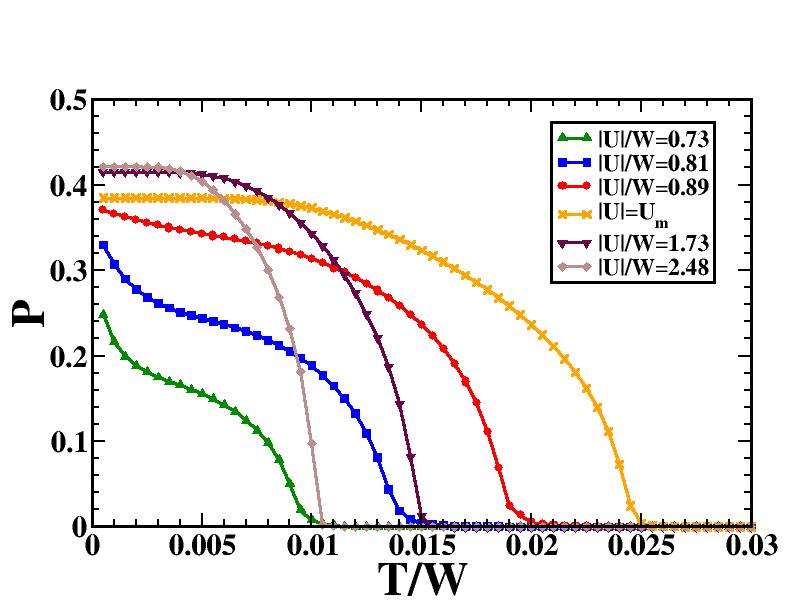}
\end{minipage}}
\hspace{0.45cm}
\subfigure[]{
\begin{minipage}[b]{0.45\textwidth}
\includegraphics[scale=0.275]{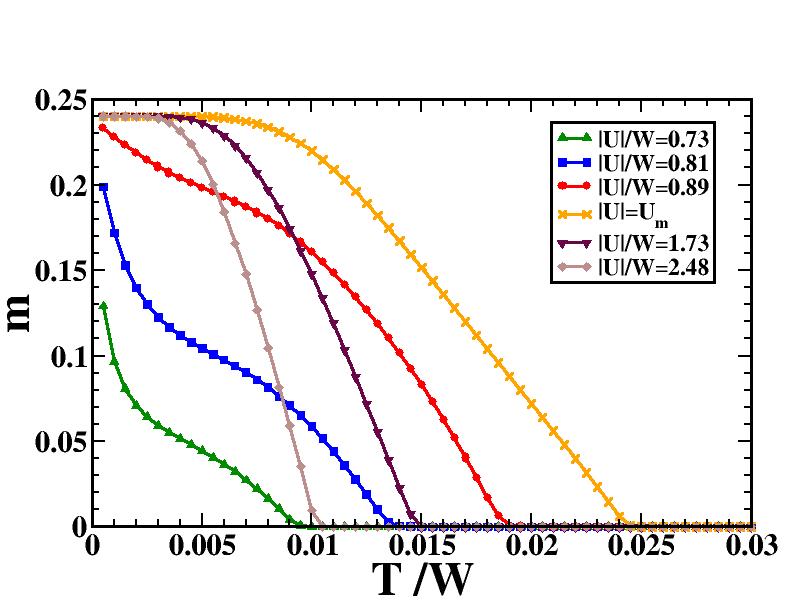}
\end{minipage}}\\
\end{center}
\caption{(a) Superconducting order parameter and (b) magnetization as a function of temperature for different values of the interaction strength $U$.
(Constrained case, $V\simeq 80W$)}
\label{SC_m_vs_T_constraint_n48}
\end{figure}
%%%%%%%%%%%%%%%%%%%%%%%%%%%%%%%%%%%%%%%%%%%%%%%%%%%%%%%%%%%%%%%%%%%%%%%%%%%%%%%%%%%%%%%%%%%%%%%%%%%%%%%%

%%%%%%%%%%%%%%%%%%%%%%%%%%%%%%%%%%%%%%%%%%%%%%%%%%%%%%%%%%%%%%%%%%%%%%%%%%%%%%%%%%%%%%%%%%%%%%%%%%%%%%%%
\begin{figure}
\begin{center}
\includegraphics[scale=0.275]{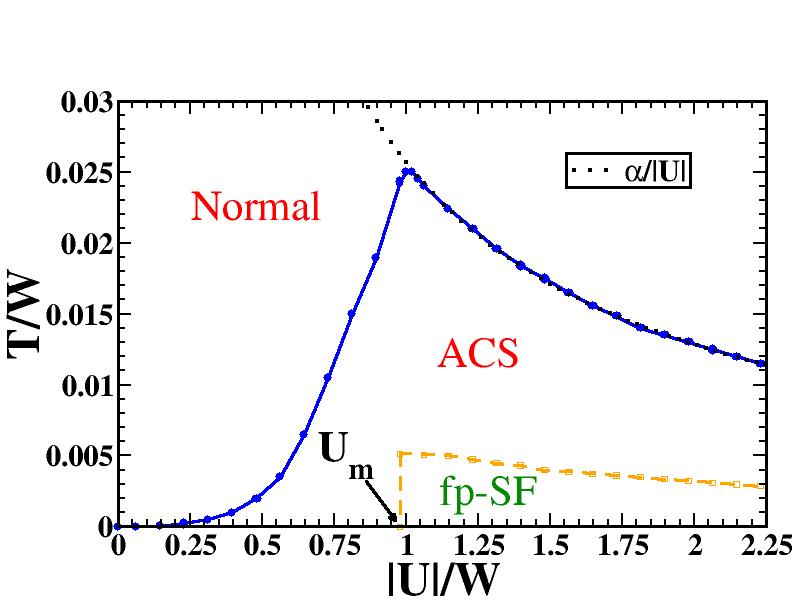}  
\end{center}
\vspace{-0.4cm}
\caption{(Color online) Phase diagram of the model on the cubic lattice with three-body constraint. The solid blue line separates normal and color
superfluid phases. Below the dashed orange line the system is fully polarized. The dotted black line describes the strong coupling behavior of the critical
temperature and is obtained by a fitting procedure.}
\label{PD_constraint_n0.48}
\end{figure}
%%%%%%%%%%%%%%%%%%%%%%%%%%%%%%%%%%%%%%%%%%%%%%%%%%%%%%%%%%%%%%%%%%%%%%%%%%%%%%%%%%%%%%%%%%%%%%%%%%%%%%%%

We also investigated finite-temperatures properties for the three-dimensional case using DMFT.
In Fig. \ref{SC_m_vs_T_constraint_n48}, we show the evolution at finite temperature $T$ of the SF order 
parameter $P$ and of the magnetization $m$  at fixed values of the interaction $U$. At low temperatures,
the system is superfluid and the magnetization finite. With increase of the temperature, both $P$ 
and $m$ decrease and then vanish \emph{simultaneously} at the critical temperature $T=T_c(U)$.
This clearly reflects the close connection between superfluid properties and magnetism in the 
$SU(3)$-symmetric case and is markedly different from the strongly
asymmetric case which we studied in Ref. \cite{asymm_short}, where the density imbalance survives
well above the critical temperature. 

It is however remarkable that for $|U| > U_m \approx W$, $m(T)$ and $P(T)$ clearly show in Fig. 
\ref{SC_m_vs_T_constraint_n48} the existence of a plateau at finite $T$, indicating that the system
 stays in practice fully polarized in a finite range of temperatures. This allows us to define 
operatively a second temperature $T_p(U)$ below which the system is fully polarized, while for 
$T > T_p$ instead the magnetization decreases and eventually vanishes at $T_c$. 

We summarize these results in the phase diagram in Fig. \ref{PD_constraint_n0.48}. Inside the region marked in orange
($|U| > U_m$ and $T<T_p$) the system is fully polarized and therefore identical to the $SU(2)$ superfluid case.
As we will see in the next section, in a canonical ensemble where the total number of particles $N_\sigma$
of each species is fixed, this analogy is not any more correct and we have to invoke the presence of domain formation
to reconcile these findings with the global number conservation in each channel. Outside this region and below
$T_c$ (solid blue line in Fig.  \ref{PD_constraint_n0.48}), the c-SF is partially magnetized and therefore 
intrinsically different from the case with only two species. This
is also visible in the behavior of the critical temperature where the $SU(3)$-symmetry is
restored in the normal phase.
We found indeed that the critical temperature first increases with the interaction strength $|U|$, similarly
to the $SU(2)$ case. Then for $|U|=U_m$, the critical temperature $T_c$ suddenly changes trend and for larger
$|U|$ a power-law decrease $T_c \propto 1/|U|$ occurs as shown in Fig. \ref{PD_constraint_n0.48}. In the
$SU(2)$ symmetric case this power-law behavior only appears for very large $|U|$ (bosonic limit) \cite{toschi},
while in the $SU(3)$ case this regime occurs \emph{immediately} for $|U|> U_m$. The smooth crossover in $T_c(U)$
and the maximum of in the critical temperature characteristic of the $SU(2)$
case, here are replaced by a cusp at $|U|=U_m$, which marks the abrupt transition from one regime to the other. 

\section{Domain Formation}
\label{Phase_Separation}

One of the main results of this work is the close connection between superfluidity and magnetization in the c-SF phase. 
Indeed we found that in the c-SF phase, away from the particle-hole symmetric point, the magnetization is always non-zero.
On the other hand ultracold gas experiments are usually performed under conditions where the global number of particles 
$N_\sigma=\sum_i n_{i,\sigma}$ in each hyperfine state is conserved, provided spin flip processes are
suppressed. The aim of this section is to show that domain formation provides a way to reconcile our findings with these
circumstances. In particular, combining DMFT and VMC findings, we will show that a globally homogeneous c-SF phase is unstable
with respect to formation of domains with different c-SF phases in the thermodynamic limit.

To be more specific, we will consider the case when the global numbers of particles in each species are the same, i.e.
$N_1=N_2=N_3=N/3$, at $T=0$, though the discussion can be easily generalized to other cases. The simplest solution 
compatible with $N_\sigma=N/3$ is clearly a non-polarized c-SF phase with energy $E_{hom}$ per lattice site. 
This phase is actually unstable and therefore not accessible in a grand canonical approach like DMFT, where we fix the
global chemical potential $\mu$ and calculate the particle densities $n_\sigma$ as an output. Since, as shown in 
Sec. \ref{unconstrained} and Sec. \ref{constrained}, the system is \emph{spontaneously} magnetized in the color superfluid phase
out of half-filling, there is no way to reconcile the DMFT result with the global constraint $N_\sigma=N/3$ assuming the presence
of a single homogeneous phase.      
The VMC approach, on the other hand, operates in the canonical ensemble, and it can be used to
estimate the ground state energy per lattice site for specific trial configurations. For the homogeneous
configuration, we have $E_{hom}=E(m = 0)_n$, where $n=N/M$
and $M$ is the number of lattice sites.

%%%%%%%%%%%%%%%%%%%%%%%%%%%%%%%%%%%%%%%%%%%%%%%%%%%%%%%%%%%%%%%%%%%%%%%%%%%%%%%%%%%%%%%%%%%%%%%%%%%%%%%%
\begin{figure}
\begin{center}
\subfigure[]{
\begin{minipage}[b]{0.3\textwidth}
\includegraphics[scale=0.15]{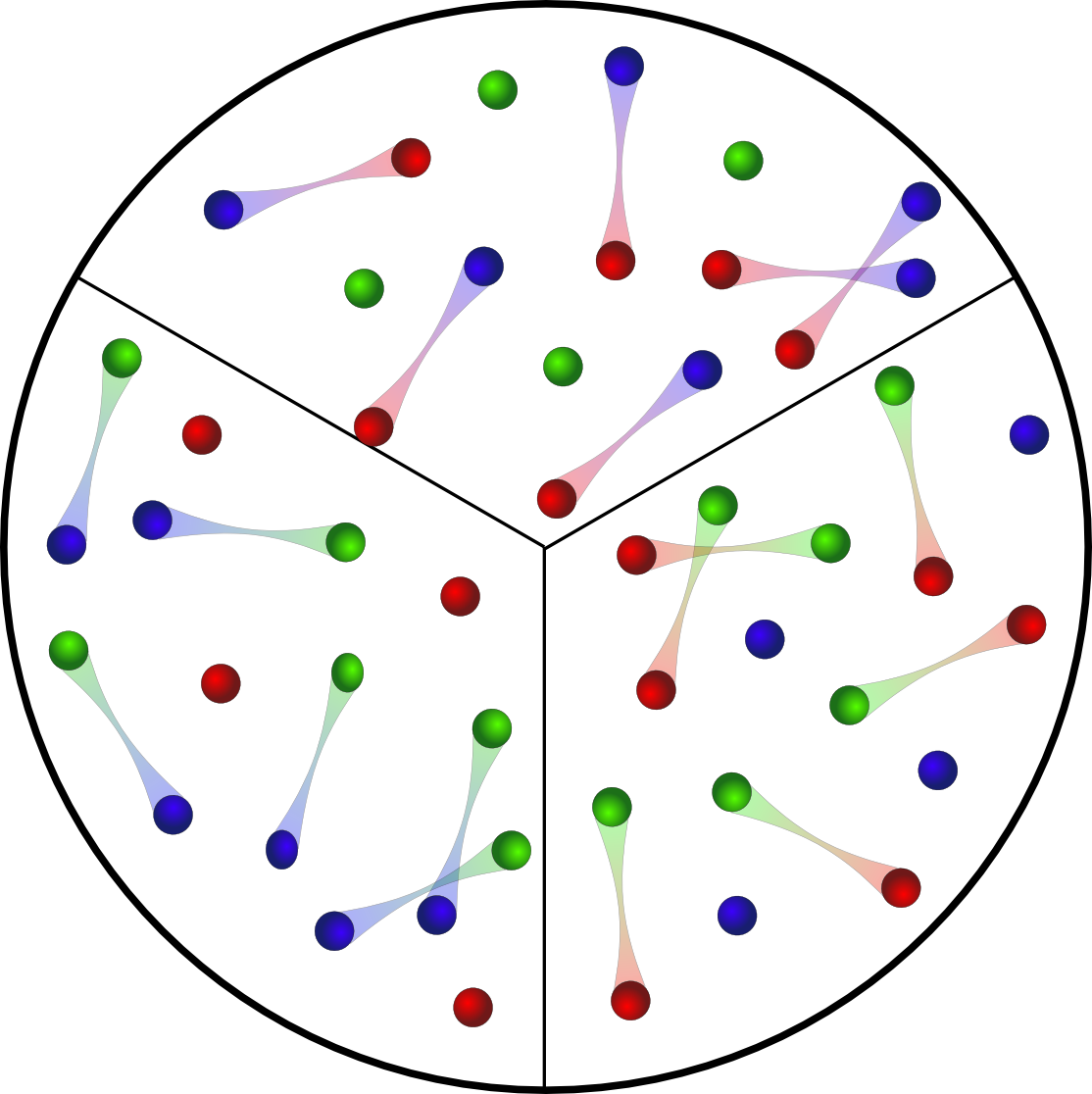}
\end{minipage}}
\hspace{0.25cm}
\subfigure[]{
\begin{minipage}[b]{0.3\textwidth}
\includegraphics[scale=0.15]{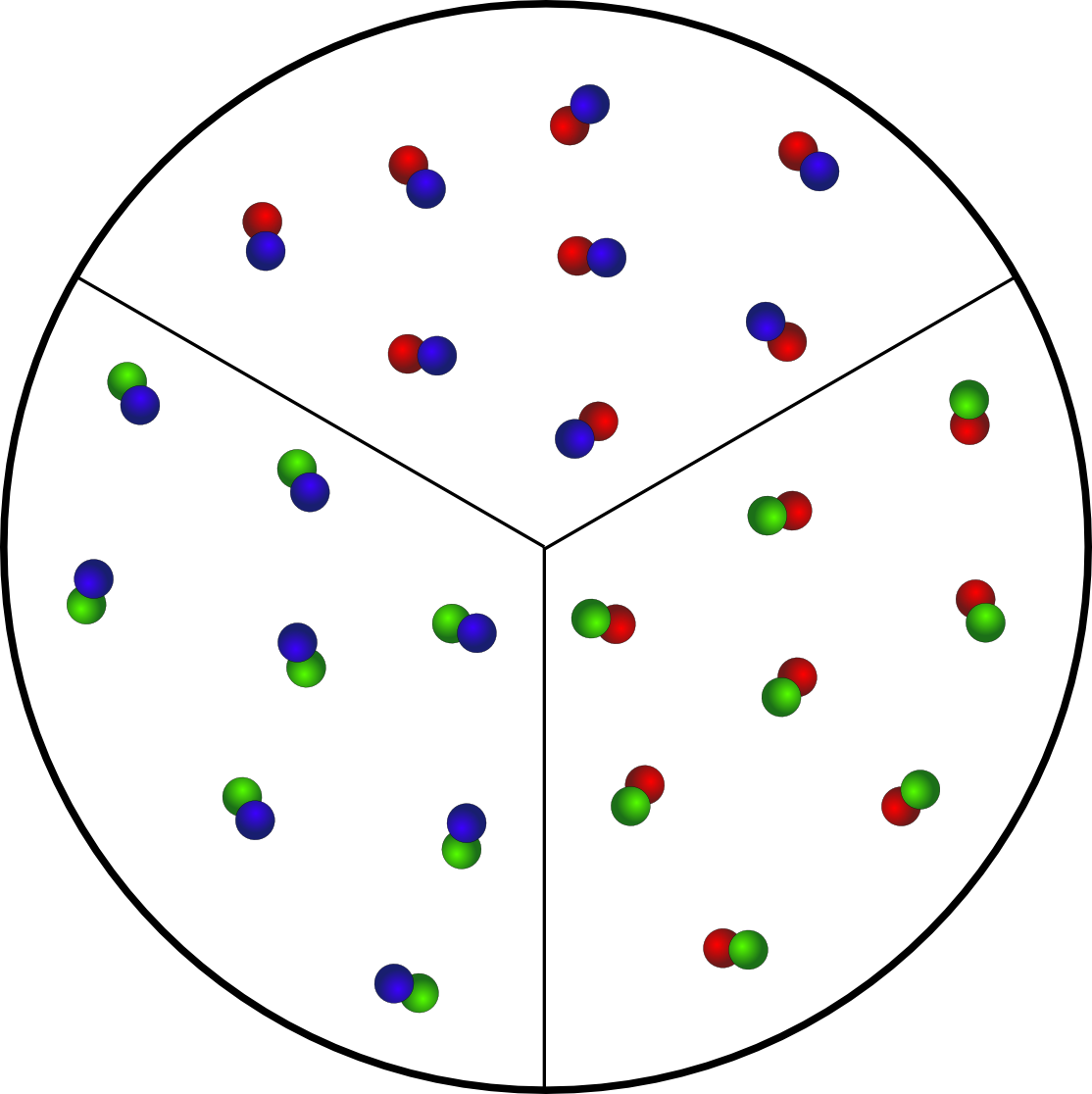}
\end{minipage}}
\hspace{0.25cm}
\subfigure[]{
\begin{minipage}[b]{0.3\textwidth}
\includegraphics[scale=0.15]{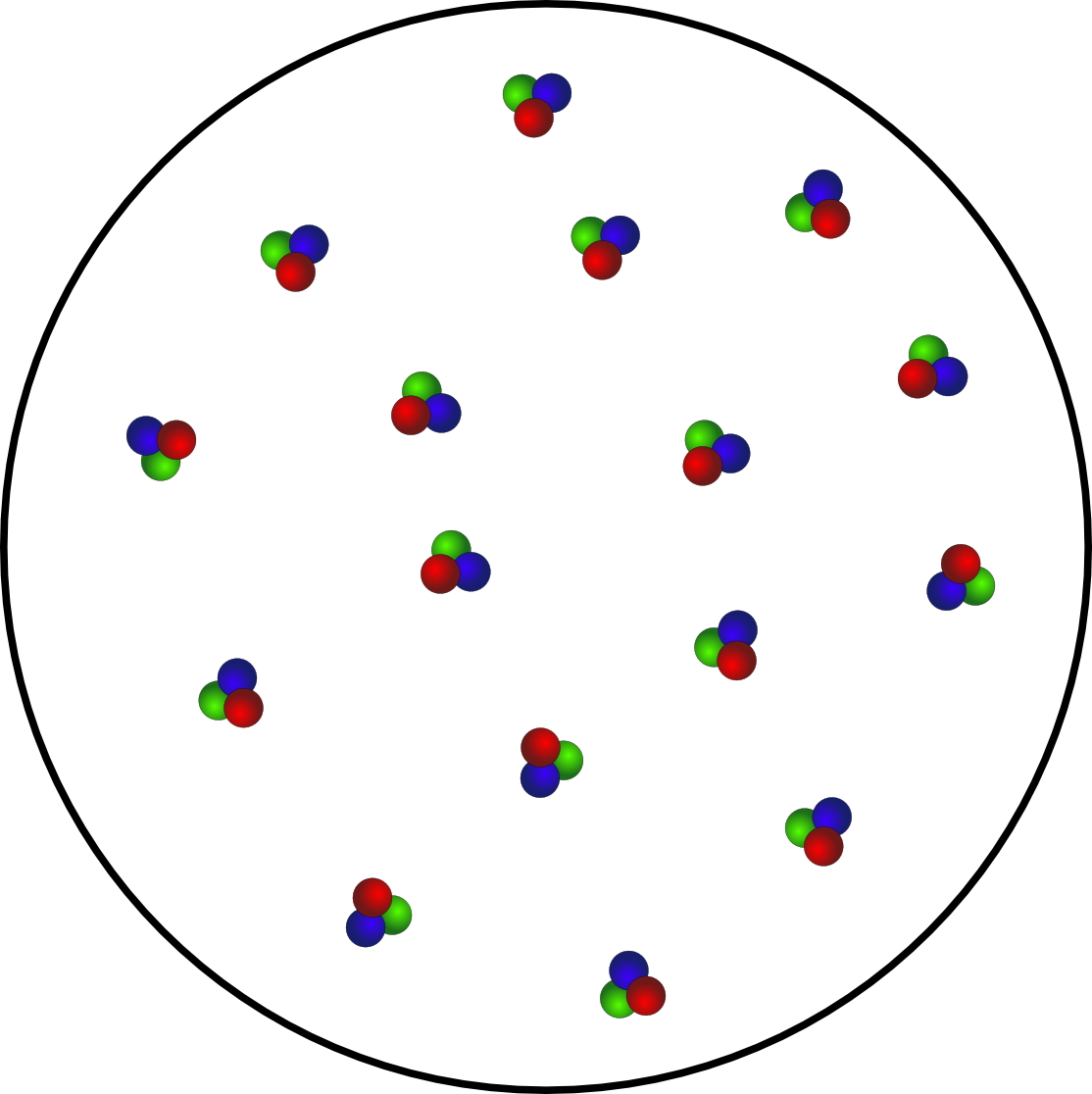}
\end{minipage}}\\
\end{center}
\vspace{-0.4cm}
\caption{(Color online) Schematic picture of the phases of a $SU(3)$-symmetric mixture of three-species fermions for the total particle
numbers in each species $N_\sigma = N/3$ away from half filling. (a) visualizes the ground state configuration at weak to
intermediate coupling in both the unconstrained ($V=0$) and the constrained ($V=\infty$) case. Irrespective of the presence of the constraint, a finite magnetization
points at domain formation in experiments with fixed $N_\sigma$ (see text); a specific example of a phase-separated configuration is plotted. Increasing the
attraction strength reveals substantial differences between the two cases: (b) In the constrained case, domain formation persists to strong coupling, in parallel to
the 3-component asymmetric situation \cite{asymm_short}. The unpaired species are expelled from the paired regions, pairing up in
other spatial domains. (c) In the unconstrained case instead, a spatially homogeneous trionic phase emerges \cite{hof3,hof4}.}
\label{PS}
\end{figure}
%\vspace{1cm}
%%%%%%%%%%%%%%%%%%%%%%%%%%%%%%%%%%%%%%%%%%%%%%%%%%%%%%%%%%%%%%%%%%%%%%%%%%%%%%%%%%%%%%%%%%%%%%%%%%%%%%%%

Let us now contrast this situation with the spatially non-uniform scenario in which we have many color superfluid domains in
equilibrium. Each of these domains corresponds to one of the solutions obtained above, and therefore this phenomenon can be seen 
as a special form of phase separation. For two or more phases to be in thermodynamic equilibrium with each other at $T=0$, they 
need to have the same value of the grand potential  per lattice site $\Omega=E-\mu n$ for the same given value of the chemical 
potential $\mu$, while the onsite density of particles for each species $n_\sigma$ can be
different in the different phases.

Possible candidate phases for the system considered in this paper are suggested by the underlying $SU(3)$-symmetry. Indeed if we
consider c-SF solutions corresponding to different gauge fixing, i.e. with pairing in different channels, they will have the same
total onsite density $n$ and therefore the same energy and grand potential, since they correspond to different realizations of the
spontaneously broken symmetry. If we consider for simplicity only the three solutions with pairing between the \emph{natural species}
sketched in Fig. \ref{PS}, then this mixture of phases has globally the same number of particles $N_\sigma=N/3$ 
in each hyperfine state whenever we choose the fraction of each phase in the mixture to be $\alpha=1/3$ and $n=N/M$
in each domain. In fact in each domain we have the same densities $n_p$ in the paired channel and $n_u$ for the unpaired fermions,
even though they involve different species in different domains. This scenario is therefore compatible with the global number
constraint $N_\sigma=N/3$ and we can compare its energy with the energy $E_{hom}$ of the globally homogeneous c-SF phase.
The VMC calculations reported in Fig. \ref{Energy_difference} clearly indicate that for a fixed onsite density $n$,
the ground state energy per lattice site is lower by having a finite magnetization, i.e. $E(m)_n < E(0)_n$ and therefore 
$E_{hom} > E_{phase-separated}= \alpha \sum_{i=1}^3 E_{i}=E(m)$ and $E_i=E(m)$ is the energy per lattice site in the i-th domain.
Thus a globally homogeneous c-SF phase has higher energy than a mixture of polarized domains with the same
$N_\sigma$ and is therefore unstable with respect to phase separation. 

It should be noted however, that the configuration sketched in Fig. \ref{PS} only represents the \emph{simplest possible} scenario
compatible with the global boundary conditions $N_\sigma=N/3$. Indeed in the $SU(3)$-symmetric case we have 
continuous set of equivalent solutions, since solutions obtained continuously rotating the pairing state
from 1-2 to a generic linear combination of species have the same energy and are therefore equally good
candidates for the state with domain formation. Moreover, it is well known that having a continuous symmetry breaking
is intrinsically different from the discrete  case, because of the presence of Goldstone modes \cite{hof1}. In large
but finite systems, the surface energy at the interface between domains, which is
negligible in the thermodynamic limit, will become relevant. On one hand a continuous symmetry breaking allows the system
to reduce the surface energy cost through an arbitrarily small change of the order parameter from domain to domain, pointing
toward a scenario where a large number of domains is preferable in real systems. On the other hand, when the system is finite,
increasing the number of domains decreases their extension, reducing the bulk contribution which eventually defines number
and size of the domains at equilibrium. Based on our current approaches, we cannot address the issue of what is the real domain
configuration in a finite system, neither the question if different scenarios with microscopical modulations of the SF order
parameter take place \cite{FuldeFerrell, LarkinOvchinnikov}. Similar conclusions concerning the emergence of domain formation
in the c-SF phase have been already drawn in \cite{hof3,hof4,cherng} and also in a very recent work
\cite{baym}, which addresses the same system in continuum space. 

In real experiments both finite-size effects and inhomogeneities due to the trapping potential could play an important role in the
actual realization of the presented scenario. Furthermore, as the $SU(3)$-symmetry in the cold atomic systems is not 
fundamental but arises as a consequence of fine-tuning of the interaction parameters, imperfections will also arise from slight 
asymmetries in these parameters. We have shown before \cite{asymm_short} that in the strongly 
asymmetric limit, phase separation is a very robust phenomenon. 
We may therefore conjecture that interaction parameter asymmetries favor this scenario.

The combination of the findings in the present paper on the $SU(3)$ case with those on the strongly-asymmetric case 
in \cite{asymm_short} suggests that phase-separation in globally \emph{balanced} mixtures is a quite general feature of
three-species Fermi mixtures. However, the phases involved are in general different in different setups. In the strongly-asymmetric
case in presence of a three-body constraint, the color superfluid phase undergo a spatial separation in superfluid dimers and 
unpaired fermions \cite{asymm_short}. In this case, the presence of the constraint is crucial to the phase-separation phenomenon,
as testified by its survival well above the critical temperature for the disappearance of the superfluid phase \cite{asymm_short}. In the fully
$SU(3)$ symmetric case instead, the presence of the constraint only modifies the nature of the underlying color superfluid phase 
favoring fully polarized domains at strong coupling. The formation of many \emph{equivalent} color superfluid domains can be seen
as a special case of phase separation reflecting the $SU(3)$ symmetry. In this case the phase separation phenomenon is strongly 
connected to the superfluid and magnetic properties of the color superfluid phase and it is expected to disappear at the critical 
temperature $T_c$ and for the peculiar particle-hole symmetric point at half-filling in the unconstrained case.

\section{Conclusions}\label{conclusion}

We have studied a SU(3) attractively interacting mixture of three-species fermions in a lattice with and without a three-body constraint
using dynamical mean-field theory ($D\geq 3$) and variational Monte Carlo techniques ($D=2$). We have investigated both ground state properties
of the system and the effect of finite temperature and find a rich phase diagram. 

For the unconstrained system, we found a phase transition from a color superfluid state to a trionic phase, which shows additional charge density modulation at
half-filling. The superfluid order as well as CDW disappear with increasing temperature. 

In the presence of the three-body constraint, the  ground state is always superfluid, but for strong interactions $|U|>U_m$ the system becomes fully polarized
for fixed total density $n$. It is remarkable that according to our calculations the system stays fully polarized in a range of low temperatures. For high
temperatures a transition to the non-superfluid $SU(3)$ Fermi liquid phase is found. The critical temperature has a cusp precisely at $U_m$. This
is in contrast to the $SU(2)$-symmetric case, where a smooth crossover in the critical temperature takes place. 

The c-SF phase shows an interesting interplay between superfluid and magnetic properties. Except in the special case of half-filling, the c-SF phase \emph{always}
implies a spontaneous magnetization  which leads to domain formation in balanced 3-component mixture.

\section*{Acknowledgment}
We thank S. Jochim for insightful discussions about three-component Fermi gases. AP thanks M. Capone for valuable discussions and financial support.
Work in Frankfurt is supported by the German Science Foundation DFG through Sonderforschungsbereich
SFB-TRR 49. Work in Innsbruck is supported by the Austrian Science Fund through SFB F40 FOQUS and EUROQUAM\_DQS (I118-N16). SYC also acknowledges support from ARO
W911NF-08-1-0338 and NSF-DMR 0706203. This research was supported in part by the National Science Foundation under Grant No. PHY05-51164.

\appendix
\section{Derivation of the strong coupling Hamiltonians}\label{Appendix}

\subsection{Constrained case}

In order to derive a perturbative strong-coupling Hamiltonian for the constrained case we make use of the Wolff-Schrieffer transformation \cite{macdonald88}
\begin{equation}
{\cal H}_{pert} = {\cal P}_D e^{i{\cal S}} {\cal H} e^{-i{\cal S}} {\cal P}_D 
\end{equation}
and keep terms up to the second order in $J/U_{\sigma\sigma^\prime}$. In the expression above, ${\cal P}_D$ is the projection operator to the Hilbert subspace with
fixed numbers of double occupancies in each channel ($N_{d}^{12}$, $N_{d}^{23}$,$N_{d}^{13}$), and $e^{i{\cal S}}$ is a unitary transformation defined below.
The kinetic energy operator can be split in several contributions, where the subscripts indicate the change in the total number of double occupancies ($N_{d,0}
=  N_{d}^{12} + N_{d}^{23} + N_{d}^{13}$), i.e. 
\begin{eqnarray}
{\cal K}_0  & =& -J \sum\limits_{\langle i,j \rangle\sigma} h_{i,\bar{\sigma}} h_{i,\bar{\bar{\sigma}}} c^\dagger_{i,\sigma} c_{j,\sigma} h_{j,\bar{\sigma}}
h_{j,\bar{\bar{\sigma}}} \nonumber \\
&&-J \sum\limits_{\langle i,j \rangle\sigma} (n_{i,\bar{\sigma}}h_{i,\bar{\bar{\sigma}}}+h_{i,\bar{\sigma}}n_{i,\bar{\bar{\sigma}}} )  c^\dagger_{i,\sigma}
c_{j,\sigma} (n_{j,\bar{\sigma}}h_{j,\bar{\bar{\sigma}}}+h_{j,\bar{\sigma}}n_{j,\bar{\bar{\sigma}}} ),  \\
{\cal K}_1  & = & -J \sum\limits_{\langle i,j \rangle\sigma} (n_{i,\bar{\sigma}}h_{i,\bar{\bar{\sigma}}}+h_{i,\bar{\sigma}}n_{i,\bar{\bar{\sigma}}} ) 
c^\dagger_{i,\sigma} c_{j,\sigma} h_{j,\bar{\sigma}} h_{j,\bar{\bar{\sigma}}} ,\\ 
{\cal K}_{-1} & =& -J \sum\limits_{\langle i,j \rangle\sigma} h_{i,\bar{\sigma}}
h_{i,\bar{\bar{\sigma}}}
c^\dagger_{i,\sigma} c_{j,\sigma}(n_{j,\bar{\sigma}}h_{j,\bar{\bar{\sigma}}}+h_{j,\bar{\sigma}}n_{j,\bar{\bar{\sigma}}} ) \, .
\end{eqnarray}
Here $n_{i\sigma}=c_{i,\sigma}^\dagger c_{i,\sigma}^{\phantom\dagger}$, $h_{i,\sigma}=1-n_{i\sigma}$ and $\sigma \not= \bar\sigma \not= \bar{\bar\sigma}
\not=\sigma$. 

We note that whereas ${\cal K}_0$ preserves the total double occupancy $N_{d,0}$, it contains two different types of terms: (i) terms that also preserve double
occupancy in each channel $N^{\sigma\sigma^\prime}_d$ ( ${\cal K}_0^a$ part) and (ii) terms that change the double occupancy in two different channels such that the 
total double occupancy stays unchanged (${\cal K}_{0}^b$ part). Thus, we can
write 
\begin{equation}
{\cal K}_0 =  {\cal K}_0^a  + {\cal K}_0^b.
\end{equation}
We can also decompose the operators that change the total number of double occupancies into
\begin{eqnarray}
{\cal K}_1 & = &  {\cal K}_1^{12}  +{\cal K}_1^{23}  +{\cal K}_1^{13} \, ,\\
{\cal K}_{-1} & = &  {\cal K}_{-1}^{12}  +{\cal K}_{-1}^{23}  +{\cal K}_{-1}^{13} \, , 
\end{eqnarray}
where the superscripts give the type of double occupancies that are being created or destroyed. The canonical transformation can be written as an expansion to the
second order
\begin{equation}
{\cal H}_{pert} = {\cal P}_D \left\{ {\cal H} + [i{\cal S},{\cal H}] + \frac{1}{2} [i{\cal S},[i{\cal S},{\cal H}]]  \right\}   {\cal P}_D,
\label{eqn_hexp}
\end{equation}
where ${\cal H} = {\cal K}_0 + {\cal K}_1 + {\cal K}_{-1} + {\cal V}$ and 
${\cal V} = \sum\limits_{i,\sigma<\sigma'} U_{\sigma\sigma^\prime} n_{i,\sigma} n_{i,\sigma'}$.
Then, we choose
\begin{equation}
i{\cal S} = \sum\limits_{\sigma < \sigma'} \left\{ \frac{1}{U_{\sigma\sigma'}} ({\cal K}^{\sigma \sigma'}_1 - {\cal K}^{\sigma \sigma'}_{-1}) +
\frac{1}{(U_{\sigma \sigma'})^2} \left( [{\cal K}^{\sigma \sigma'}_1, {\cal K}_0] 
+ [{\cal K}^{\sigma \sigma'}_{-1}, {\cal K}_0] \right) \right\} \, .
\label{eqn_is2}
\end{equation}
Inserting Eq. (\ref{eqn_is2}) into the Eq. (\ref{eqn_hexp}) we obtain
\begin{eqnarray}
&&\hspace{-2.5cm}
{\cal H}_{pert} =  {\cal V} + {\cal K}_0^a  \\
&&\hspace{-2.5cm}+
\sum\limits_{\sigma < \sigma'} \sum\limits_{\sigma'' <\sigma'''}
\frac{1}{2U_{\sigma\sigma'} U_{\sigma'' \sigma'''}} {\cal P}_D \left[({\cal K}^{\sigma \sigma'}_1-
{\cal K}^{\sigma \sigma'}_{-1}), 
[{\cal K}^{\sigma'' \sigma'''}_1,{\cal V}]  
- [{\cal K}^{\sigma'' \sigma'''}_{-1},{\cal V}]\right]{\cal P}_D
+{\cal O}(\frac{J^3}{U^2}) .\nonumber
\end{eqnarray}
Using the relation $[{\cal V}, {\cal K}_{\pm 1}^{\sigma\sigma'}] = 
\pm U_{\sigma, \sigma'} {\cal K}_m^{\sigma\sigma'}$ and applying the projection ${\cal P}_D$,  we arrive at
\begin{equation}
{\cal H}_{pert} = {\cal V} + {\cal K}_0^a + \sum\limits_{\sigma < \sigma'}
\frac{1}{U_{\sigma \sigma'}} [{\cal K}^{\sigma \sigma'}_{-1},{\cal K}^{\sigma \sigma'}_{1}] +{\cal O}(\frac{J^3}{U_{\sigma \sigma'}^2}) .
\label{eqn_pert0_A}
\end{equation}
Notice that most of the terms in the commutator become zero leaving only the 
{\it correlated hopping} terms. 

In order to write Eq. (\ref{eqn_pert0_A}) in a more practical way, we can define double occupancy operators as $d_{i,\sigma\sigma'}^\dagger \equiv c_{i,\sigma}^\dagger
n_{i,\sigma'}  h_{i,\sigma''}$ and single occupancy operators as $f_{i,\sigma}^\dagger = h_{i,\sigma'}h_{i,\sigma''} c_{i,\sigma}^\dagger$ with
$\sigma \ne \sigma' \ne \sigma'' \ne \sigma$.  With this notation, the perturbative Hamiltonian becomes 
\begin{eqnarray}
{\cal H}_{pert}  & = &  -J\sum\limits_{\langle i,j \rangle\sigma} f_{i,\sigma}^\dagger f_{j,\sigma} 
- J^2 \sum\limits_{\langle j,i \rangle; \langle i,j \rangle;\sigma < \sigma'} \frac{1}{U_{\sigma\sigma^\prime}} d^\dagger_{j,\sigma\sigma'} f_{i,\sigma}
f^\dagger_{i,\sigma} d_{j,\sigma\sigma'} \nonumber \\
&&- J^2 \sum\limits_{\langle i,j \rangle; \langle i,j \rangle;\sigma < \sigma'} \frac{1}{U_{\sigma\sigma^\prime}} d^\dagger_{i,\sigma'\sigma} f_{j,\sigma'}
f^\dagger_{i,\sigma} d_{j,\sigma\sigma'} 
+ {\cal V} + {\cal O}(\frac{J^3}{U_{\sigma \sigma'}^2})~.
\label{eqn_pert1_A}
\end{eqnarray}

For the case where the $SU(3)$-symmetry is restored ($U_{\sigma\sigma^\prime}=U$), the perturbative Hamiltonian can be written in a compact notation 
\begin{eqnarray}
\label{eqn_pert2_A}
&&\hspace{-2cm}{\cal H}_{pert} ={\cal V} -J\sum\limits_{\langle i,j \rangle\sigma} \left[ f_{i,\sigma}^\dagger f_{j,\sigma} +  d_{i,\sigma}^\dagger d_{j,\sigma}
\right] 
- \frac{J^2}{U} \sum\limits_{\langle i',i \rangle; \langle i,j \rangle;\sigma} d^\dagger_{i',\sigma} f_{i,\sigma} f^\dagger_{i,\sigma} d_{j,\sigma} \\
&& \hspace{-1.25cm}
- \frac{J^2}{U} \sum\limits_{\langle i,j' \rangle; \langle i,j \rangle;\sigma'\ne\sigma} d^\dagger_{i,\sigma'} f_{j',\sigma'} f^\dagger_{i,\sigma} d_{j,\sigma} 
+ \frac{J^2}{U} \sum\limits_{\langle i',i \rangle\sigma'; \langle i,j \rangle\sigma} f^\dagger_{i',\sigma'} d_{i,\sigma'} d^\dagger_{i,\sigma} f_{j,\sigma} 
+  {\cal O}(\frac{J^3}{U^2}) ,\nonumber
\end{eqnarray}
where the double occupancy operator is now defined as $d_{i,\sigma}^\dagger = c_{i,\sigma}^\dagger  (h_{i,\sigma'}  n_{i,\sigma''}
+h_{i,\sigma''}  n_{i,\sigma'})$.

\subsection{Unconstrained case}

Without the 3-body constraint three fermions with different hyperfine states can occupy the same lattice site and we expect them to form trionic bound
states at sufficiently strong coupling. 

According to perturbation theory up to third order we could have two different contributions:  (i) one of the fermions hops to one of the neighboring sites and
returns back to the original site (second order perturbation), (ii) all three  fermions hop to the same nearest neighbor site (third order perturbation).  As
we show below, due to the first process there is an effective interaction between trions on nearest neighbor sites. Also due to this process the onsite energy
has to be renormalized. The second process (ii) describes the hopping of a local trion to a neighboring site.

The energy gain due to virtual processes, when one of the fermions is hopping to a nearest neighboring site and returning back, can be easily determined within
second-order perturbation theory
\begin{equation}
\Delta E ={\sum_{i,\sigma}}^\prime\frac{|\langle i \sigma|{\cal H}|t_0\rangle|^2}{E_{t_0}-E_{i\sigma}},
\end{equation}
where $\sum_{i}^\prime$ denotes summation only over the nearest neighbors of the trion. Here $|t_0\rangle$ describes a local trionic state at lattice site $0$,
while by $|i\sigma\rangle$ we define a state where site $i$  is occupied by a fermion with spin $\sigma$, while two other fermions stay in the lattice site $0$. 
One can easily calculate that  $|\langle i \sigma|{\cal H}|t_0\rangle|^2=J^2$ and $E_{t_0}-E_{i\sigma}=U_{\sigma \sigma'}+U_{\sigma \sigma''}$, where
$\sigma\not=\sigma'\not=\sigma'' \not=\sigma$. So we obtain 
\begin{equation}
\Delta E=\frac{zJ^2}{U_{12}+U_{13}}+ \frac{zJ^2}{U_{12}+U_{23}}+ \frac{zJ^2}{U_{13}+U_{23}} \, ,
\end{equation}
where $z$ is the number of the nearest neighbor lattice sites. 

The calculation above assumes that neighboring sites of a trion are not occupied. If one of the neighboring sites is occupied by another trion, then the energy gain
per trion is given by
\begin{equation}
\Delta E_1= \frac{(z-1)J^2}{U_{12}+U_{13}}+ \frac{(z-1)J^2}{U_{12}+U_{23}}+ \frac{(z-1)J^2}{U_{13}+U_{23}} \, .
\end{equation}
The effective interaction between two trions on neighboring sites is therefore
\begin{equation}
V_{eff}=\Delta E_1-\Delta E_0= -\left(\frac{J^2}{U_{12}+U_{13}}+ \frac{J^2}{U_{12}+U_{23}}+ \frac{J^2}{U_{13}+U_{23}}\right) \, .
\end{equation}
For the $SU(3)$-symmetric case this expression is simplified and we obtain
\begin{equation}
V_{eff}= -\frac{3J^2}{2U}=\frac{3J^2}{2|U|} \,.
\end{equation}
Therefore the nearest neighbor interaction between trions is repulsive in the $SU(3)$-symmetric case. 

The next step is to calculate the effective hopping of the trions. For this purpose one has to use third order perturbation theory
\begin{equation}
-J_{eff}=\sum_{\sigma,\sigma^\prime}^{\sigma\not=\sigma^\prime}\frac{\langle t_0|{\cal H}|\sigma\rangle\langle\sigma|{\cal H}|\sigma\sigma^\prime\rangle
\langle \sigma\sigma^\prime|{\cal H}|t_1\rangle}{(E_0-E_\sigma)(E_1-E_{\sigma\sigma^\prime})} .
\end{equation}
Here $|t_0\rangle$ and $|t_1\rangle$ define local trions on lattice site $0$ and the neighboring lattice site $1$ respectively, $|\sigma\rangle$ defines a state
where a fermion with spin $\sigma$ occupies the lattice site $1$, and two other fermions are occupying the lattice site $0$. Conversely $|\sigma\sigma^\prime\rangle$
defines a state where two fermions with spins $\sigma$ and $\sigma^\prime$ occupy the lattice site $1$. On the lattice site $0$ we have only a fermion with spin
$\sigma^{\prime\prime}\not=\sigma,\sigma^\prime$. For any $\sigma$ and $\sigma^\prime$ the matrix elements are given by $\langle t_0|{\cal
H}|\sigma\rangle=\langle\sigma|{\cal H}|\sigma\sigma^\prime\rangle =\langle \sigma\sigma^\prime|{\cal H}|t_1\rangle=-J $,  $E_{t_0}-E_\sigma
=U_{\sigma\sigma'}+U_{\sigma\sigma''}$ and $E_{t_1}-E_{\sigma\sigma^\prime}=U_{\sigma \sigma''}+U_{\sigma'\sigma''}$, where $\sigma$, $\sigma'$
and $\sigma''$ are three different hyperfine-spins. 

So we obtain
\begin{equation}
J_{eff}=\sum_{\sigma,\sigma^\prime}^{\sigma \not=\sigma^\prime}\frac{J^3}{(U_{\sigma\sigma'}+U_{\sigma\sigma''}) (U_{\sigma\sigma''}+U_{\sigma'\sigma''})} .
\end{equation}
where $\sigma$, $\sigma'$ and $\sigma''$ are different from each other in the sum.

In the $SU(3)$-symmetric case, the expression again simplifies to
\begin{equation}
J_{eff}=\frac{3J^3}{2U^2}.
\end{equation}
So we obtain the following effective Hamiltonian \cite{csaba}
\begin{equation}
\label{trion_Hamiltonian_A}
{\cal H}_{eff}=-J_{eff} \sum_{\langle i,j\rangle}t_i^\dagger t_j + 
V_{eff}\sum_{\langle i,j\rangle} n_i^T n_j^T \, .
\end{equation}
Here $t_i^\dagger$ is the creation operator of a local trion at lattice site $i$ and $n_{i}^T =t_i^\dagger t_i$ is the trionic number operator.

\end{document}